\documentclass[12pt]{article}
\usepackage{mathpazo}
\usepackage[pagewise]{lineno}
\usepackage{amssymb}
\usepackage{graphicx}
\usepackage{amsfonts}
\usepackage{amsmath}
\usepackage{amsthm}
\usepackage{mathtools}
\usepackage{ascmac}
\usepackage{bm}
\usepackage{color}
\usepackage{pgfplots}
\pgfplotsset{compat=1.16}
\usepackage{here}
\usepackage{multirow}
\usepackage[hang,small,bf]{caption}
\usepackage[subrefformat=parens]{subcaption}
\usepackage{authblk}
\usepackage[colorlinks=true, citecolor=teal, linkcolor=teal, urlcolor=teal]{hyperref}
\usepackage[authoryear]{natbib}
\usepackage{threeparttable}
\usepackage{comment}
\usepackage{stackengine}
\usepackage{tikz}
\usetikzlibrary{positioning, calc}
\usetikzlibrary{arrows.meta}
\usetikzlibrary{bending}
\usepackage{setspace}

\def\E#1{\mathbb{E}\left[#1\right]}

\def\P#1{\mathbb{P}\left[#1\right]}

\newcommand{\indep}{\perp\!\!\!\!\perp}

\numberwithin{equation}{section}

\newtheorem{theorem}{Theorem}[section]
\newtheorem{assumption}{Assumption}[section]

\newtheorem{example}{Example}

\newtheorem{prop}{Proposition}[section]
  
\theoremstyle{definition}
\newtheorem{remarkx}{Remark}
 \newenvironment{remark}
  {\pushQED{\qed}\remarkx}
  {\popQED\endremarkx}

\makeatletter
\renewenvironment{proof}[1][\proofname]{%
  \par\pushQED{\qed}\normalfont%
  \topsep6\p@\@plus6\p@\relax
  \trivlist\item[\hskip\labelsep\bfseries#1\@addpunct{.}]%
  \ignorespaces
}{%
  \popQED\endtrivlist\@endpefalse
}
\makeatother

\makeatletter
\def\blfootnote#1{%
  \begingroup
  \renewcommand\thefootnote{}\footnote{#1}%
  \addtocounter{footnote}{-1}%
  \endgroup
}
\makeatother

\makeatletter
\renewcommand*{\@fnsymbol}[1]{\ensuremath{\ifcase#1\or \flat\or * \else\@ctrerr\fi}}
\makeatother

\renewcommand{\footnotesize}{\fontsize{10pt}{20pt}\selectfont}

\title{\bf Bounds on Extrapolated Treatment Effects in Regression Discontinuity Designs}
\author[$\sharp$]{Yuta Okamoto}
\author[$\flat$]{Yuuki Ozaki}
\affil[$\sharp$]{Graduate School of Economics, Hitotsubashi University}
\affil[$\flat$]{Attax Co., Ltd.}
\date{}

\begin{document}
\maketitle
\doublespacing
\blfootnote{${}^{\sharp}$\href{mailto:yuta.okamoto@r.hit-u.ac.jp}{yuta.okamoto@r.hit-u.ac.jp} (corresponding author). 
Part of this work was conducted while we were Ph.D. students at Kyoto University. An earlier version of this manuscript received the 2025 Kanematsu Prize from the Research Institute for Economics and Business Administration, Kobe University. Okamoto gratefully acknowledges financial support from JST SPRING (Grant Number JPMJSP2110).}
\vspace{-2cm}
\begin{abstract}
    External validity is a central concern in regression discontinuity (RD) designs. To assess treatment effect heterogeneity and the representativeness of the conventional RD estimand, this paper studies the identifying content of monotonicity and smoothness restrictions for treatment effects away from the cutoff in both canonical and multi-cutoff RD designs. In the canonical design, monotonicity in the running variable alone has limited identifying power and generally yields only one-sided bounds. In multi-cutoff designs, by contrast, our proposed monotonicity restrictions along both the running-variable and cutoff-group dimensions yield a support-free characterization of the identified set. In both settings, smoothness restrictions may further tighten the identified set, underscoring the value of combining shape and smoothness restrictions. 
    These sharp identified sets, or their outer approximations, are amenable to bias-aware inference and moment-inequality-based inference procedures.
    An empirical application illustrates the practical usefulness of our approach.
\end{abstract}


{\textbf{Keywords:} Causal inference, external validity, heterogeneity, shape restriction}



\newpage
\section{Introduction}\label{sec:intro}
Regression discontinuity (RD) designs are among the most credible quasi-experimental methods for identifying causal effects.
In the RD framework, treatment status changes discontinuously at a known threshold, allowing for identification of the treatment effect at the cutoff, provided a mild smoothness assumption holds \citep{Hahn_etal:2001}.
This feature provides RD designs with strong internal validity.

However, this internal validity stems from the local nature of the RD designs, and hence, their external validity often remains uncertain.
This constitutes a primary limitation of RD designs, as the treatment effect at the cutoff is not necessarily the only parameter of substantive interest. To obtain a more comprehensive understanding of treatment effects, researchers may wish to infer effects at values of the running variable away from the cutoff. For example, does the treatment also benefit students whose scores exceed the admission cutoff by a substantial margin, rather than only those who barely qualify? Standard RD designs provide limited information for answering such questions, and the extrapolation of RD treatment effects therefore remains a ``crucial open question'' \citep{Abadie_Cattaneo:2018}.

Addressing the limited external validity of RD designs requires additional sources of identifying information. One promising approach is to impose monotonicity restrictions on the regression functions. Such restrictions are often supported by economic intuition and therefore provide a credible starting point for extrapolation. Another approach is to exploit the presence of multiple cutoffs, a setting frequently encountered in empirical applications \citep{Cattaneo_etal2016jop, Cattaneo_etal:2021JASA_extrapolating}. 
This paper investigates the identifying power of monotonicity restrictions under both canonical and multiple-cutoff designs, and further examines how smoothness restrictions, which are commonly imposed in the recent RD literature, tighten identification in each setting.

Our contributions are threefold. 
First, we study the identifying power of monotonicity restrictions in both canonical and multi-cutoff RD designs. 
Although monotonicity assumptions with respect to the running variable have been employed in the RD literature to improve inference (e.g., \citealp{Babii_Kumar:2023, Kwon_Kwon:2020}), their identifying content for extrapolating RD treatment effects has received little attention. 
We further introduce a group-monotonicity condition tailored to multi-cutoff RD designs. This condition requires the conditional mean function for a higher-cutoff group to lie above that for a lower-cutoff group, and may be particularly plausible when multiple cutoffs arise from group heterogeneity or institutional rules intended to reduce disparities or implement affirmative-action policies.
In multi-cutoff designs, we show that imposing monotonicity restrictions on conditional mean functions along both the running-variable and cutoff-group dimensions yields a support-free characterization of the identified set when the two monotonicity restrictions are aligned.

Second, we embed our identification analysis in a broad sensitivity-analysis framework. We combine our baseline identification results with smoothness restrictions commonly used in the RD and extrapolation literature. 
We show and empirically illustrate that monotonicity and smoothness restrictions can play complementary identifying roles: smoothness restrictions tend to be particularly informative near the cutoff, whereas monotonicity restrictions can retain identifying power farther away from the cutoff.
We also show that the interaction between monotonicity and smoothness restrictions can, in some cases, yield a tighter identified set than simply intersecting the sharp identified sets obtained under each restriction separately.

Finally, we provide a baseline specification that nests the extrapolation result under the constant-bias assumption of \cite{Cattaneo_etal:2021JASA_extrapolating}.
We also characterize the conditions under which the identified set obtained under our monotonicity assumptions contains the point-identified value implied by the constant-bias assumption.
These results make our proposed bounds well suited for sensitivity analysis or layered analyses in which they are used as a complementary approach to point identification under the constant-bias assumption.

Our framework accommodates flexible combinations of assumptions and alternative smoothness restrictions beyond those explicitly considered in this article. 
This flexibility allows empirical researchers to conduct identification analyses that reflect the institutional context and incorporate their substantive knowledge. 
We also discuss how general inference methods, such as bias-aware inference and moment-inequality-based approaches, can be applied in these more general settings.

The remainder of this article is organized as follows.
The rest of this section reviews the related literature.
Section \ref{sec:ident} presents our first main identification results under monotonicity assumptions alone.
Section \ref{sec:smooth} discusses the additional identifying content of smoothness restrictions.
Section \ref{sec:inf} discusses estimation and inference procedures, and Section \ref{sec:emp} illustrates the empirical relevance of our approach through an empirical application.
Section \ref{sec:concl} concludes.
Appendices A--D of the Online Appendix (OA) contain proofs, results for multi-cutoff RD designs with more than two cutoffs, an extension to one-sided fuzzy RD designs, and additional theoretical results.

\subsection*{Related Literature}

While most of the extensive methodological literature on RD designs focuses on internal validity (see \citealp{Cattaneo_Titiunik:2022} for a review), a growing literature studies external validity. \cite{Angrist_Rokkanen:2015jasa} exploits conditional mean independence of potential outcomes and the running variable, while \cite{Bertanha_Imbens:2020jbes} assumes conditional independence of potential outcomes and compliance types given the running variable. \cite{Dong_Lewbel:2015} and \cite{Cerulli_etal:2017} study marginal extrapolation and marginal changes in the cutoff under smoothness conditions, whereas \cite{Bertanha:2020} identifies average effects of non-marginal counterfactual policies with many cutoffs. \cite{Mehta:2019} derives bounds assuming optimal cutoff choice by an informed policymaker, and \cite{Deaner_Kwon:2025} develops an extrapolation approach using an additional covariate satisfying comonotonicity.

Most closely related to our work, \cite{Cattaneo_etal:2021JASA_extrapolating} leverages multiple cutoffs to extrapolate treatment effects. They introduce the constant-bias assumption, which requires the conditional expectation functions for untreated potential outcomes to be parallel across cutoff groups, and establish point identification.
Relatedly, \cite{zhang:2024} relaxes the constant-bias assumption by introducing a bias-bounding constant and studies the optimal cutoff-assignment problem.
\cite{Sun:2023} also relaxes the constant-bias assumption by introducing various bias-bounding conditions and studies an extrapolation problem similar to ours.

Although closely related, these studies differ from ours along several important dimensions.
First, none of them studies the identifying power of monotonicity restrictions. 
Second, under multi-cutoff designs, our identification analysis under monotonicity restrictions alone characterizes the sharp identified set without requiring the researcher to specify any sensitivity parameters, which constitutes a key distinction from the bias-bounding approaches of \cite{Sun:2023} and \cite{zhang:2024}.
Moreover, we combine the smoothness conditions considered in these studies with our proposed monotonicity restrictions, highlight the complementary identifying roles they can play, and show that their interaction can, in some cases, yield a sharp identified set that is tighter than the intersection of the identified sets obtained under each restriction separately.

At the same time, these studies have distinct strengths relative to ours. \cite{Cattaneo_etal:2021JASA_extrapolating} achieves point identification; \cite{Sun:2023} considers a broader range of smoothness restrictions than we do; and \cite{zhang:2024} establishes the optimality of its proposed cutoff-assignment rule, an issue that lies beyond the scope of this paper.
Our work thus makes a complementary contribution to the literature on RD extrapolation.

More broadly, this article is related to the causal inference literature that uses shape restrictions. In difference-in-differences (DID) settings, for example, \cite{Manski_Pepper:2018} and \cite{Rambachan_Roth:2023} propose bounded-variation-type restrictions to relax the conventional parallel trends assumption. Given the analogy between the parallel trends assumption and the constant-bias assumption, our work shares a similar motivation: replacing a potentially restrictive point-identifying assumption with more credible restrictions that yield partial identification.

\section{Identification Analysis under Monotonicity Restrictions}\label{sec:ident}


We begin by introducing a unified setup that accommodates both canonical and multi-cutoff RD designs.
Let $Y_i \in \mathbb R$ denote the outcome variable, $X_i \in \mathcal X \subseteq \mathbb R$ the running variable (RV) with support $\mathcal X$, and $D_i\in\{0,1\}$ be a treatment indicator, which takes one if $i$ is treated and zero otherwise. 
Treatment assignment follows the rule $D_i = \mathbf{1}\{X_i \geq C_i\}$, where $C_i \in \mathcal C$ denotes the cutoff position, and $\mathcal C$ is the collection of threshold points. 
Let $Y_i(d)$ denote the potential outcome under treatment status $D_i=d\in\{0,1\}$. 
Define the (potential) conditional expectations and treatment effect functions as
\begin{align*}
    \mu_{d,c}(x) &\coloneqq \E{Y_i(d) | X_i = x, C_i = c},\\
    \tau_c(x) &\coloneqq \E{Y_i(1) - Y_i(0) | X_i = x, C_i = c} = \mu_{1,c}(x)-\mu_{0,c}(x).
\end{align*}
Suppose that the cutoff group $\ell \in \mathcal C$ is our target group of interest.
In the canonical single-cutoff RD case, $\mathcal C = \{\ell\}$ and $C_i = \ell$ holds almost surely.

Throughout, we maintain the following condition which assumes the continuity of the potential mean functions:
\begin{assumption}[Continuity]\label{assumption:continuity}
    $\mu_{d,c}(x)$ is continuous in $x \in \mathcal X$ for  $d\in\{0,1\}$ and $c\in\mathcal C$.
\end{assumption}

Write the observed regression function by $\mu_{c}(x) \coloneqq \E{Y_i | X_i = x, C_i = c}$.
The conventional RD designs identify $\tau_{\ell}(\ell)$ under the continuity assumption as $\tau_{\ell}(\ell) = \mu_{1,\ell}(\ell)-\mu_{0,\ell}(\ell) = \lim_{x \downarrow \ell} \mu_{\ell}(x) - \lim_{x \uparrow \ell} \mu_{\ell}(x)$, where the second equality uses Assumption \ref{assumption:continuity}.

This article concerns the identification of average treatment effects away from the cutoff point, in particular, $\tau_{\ell}(\bar{x})$ with $\bar{x} > \ell$. 
The treatment effect $\tau_\ell(\bar{x})$ is of interest for several reasons.
Its primary use is to characterize treatment-effect heterogeneity. In many policy settings, researchers are interested not only in the effect for individuals at the cutoff but also in the effect for individuals whose running-variable values lie well above it. For example, in an educational application, one may wish to know whether a program benefits not only students who barely satisfy the eligibility criterion but also those whose scores exceed the cutoff by a substantial margin.
In other words, $\tau_\ell(\bar{x})$ helps assess whether the local RD effect is representative of the effects experienced by the broader treated population or whether the benefits of treatment are concentrated near the cutoff.
Furthermore, under an additional policy-invariance assumption that the conditional mean potential-outcome functions remain unchanged when the assignment threshold is altered, $\tau_\ell(\bar{x})$ can be interpreted as the treatment effect at the new cutoff under a counterfactual policy that moves the threshold from $\ell$ to $\bar{x}$, as considered in \cite{Dong_Lewbel:2015}.

\subsection{Single-Cutoff RD Designs}\label{subsec:single}

We start with the canonical single-cutoff case, i.e., $\mathcal{C} = \{\ell\}$.
We are interested in the average treatment effect at $\bar{x} (> \ell)$:
\begin{align*}
    \tau_{\ell}(\bar{x}) = \mu_{1,\ell}(\bar{x})-\mu_{0,\ell}(\bar{x}) 
    = \mu_{\ell}(\bar{x}) - \mu_{0,\ell}(\bar{x}),
\end{align*}
where we used that the observed regression function coincides with the treated potential-outcome regression over the treated region, so that $\mu_{1,\ell}(\bar{x}) = \mu_{\ell}(\bar{x})$; Assumption~\ref{assumption:continuity} allows us to interpret this equality pointwise.
Our goal is therefore to obtain the identified set of $\mu_{0,\ell}(\bar{x})$. 

We begin by studying the identifying power of the following monotonicity condition with respect to the running variable. 
\begin{assumption}[RV Monotonicity]\label{assumption:x-monotonicity}
    Either of the following conditions holds.
    \begin{description}
        \item[(i)] For any $c \in \mathcal C$, $\mu_{0,c}(x)$ is weakly increasing in $x \in [c,\infty)\cap\mathcal X$.
        \item[(ii)] For any $c \in \mathcal C$, $\mu_{0,c}(x)$ is weakly decreasing in $x \in [c,\infty)\cap\mathcal X$.
    \end{description}
\end{assumption}
In the single-cutoff case, this condition says that the regression function $\mu_{0,\ell}(x)$ is weakly monotone with respect to the running variable over $x \ge \ell$.
The former version of the \textit{RV-monotonicity assumption} posits that the untreated potential outcome at a larger score value is weakly greater than that at a lower level.
Such a monotonicity assumption is commonly assumed in the partial identification literature (e.g., \citealp{Manski_Pepper:2000}), as well as RD literature (\citealp{Babii_Kumar:2023, Kwon_Kwon:2020}), although these papers do not consider the extrapolation issue.

The RV-monotonicity assumption is particularly plausible in many empirical RD settings. As \cite{Babii_Kumar:2023} wrote, ``[r]egression discontinuity designs encountered in empirical practice are frequently monotone."
To illustrate, consider typical educational settings:
For example, when $X_i$ represents a test score and $Y_i(0)$ denotes future earnings in the absence of any treatment, it is natural to assume that $\mu_{0,\ell}(x)$ is increasing, consistent with human capital theory. 
A similar logic applies when $X_i$ is family income level: higher-income families may be associated with greater expected earnings due to inherited ability or increased investment in human capital (e.g., \citealp{Bjorklund_etal:2006}).

\begin{figure}[t]
    \centering
    \begin{subfigure}[b]{0.49\textwidth}
        \centering
        \includegraphics[width=0.8\linewidth]{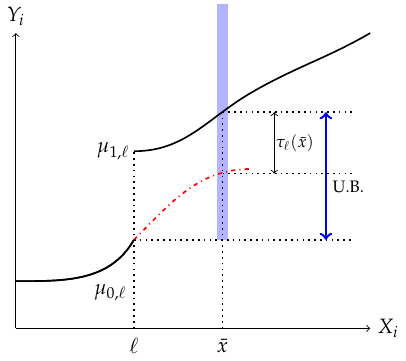}
        \caption{RV Monotonicity}
        \label{fig:tikz1}
    \end{subfigure}
    \hfill
    \begin{subfigure}[b]{0.49\textwidth}
        \centering
        \includegraphics[width=0.8\linewidth]{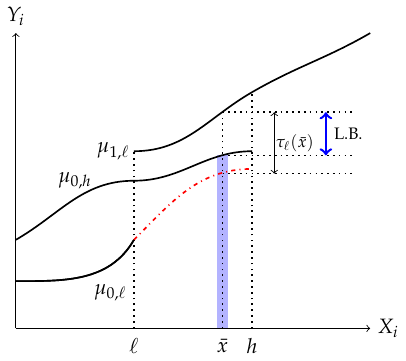}
        \caption{Group Monotonicity}
        \label{fig:tikz2}
    \end{subfigure}
    \caption{Bounds under Monotonicity Assumptions}
    \label{fig:mon}

    \begin{flushleft}
    \footnotesize
    \renewcommand{\baselineskip}{11pt}
    \textbf{Note:} Panel (a) illustrates the upper bound under Assumption \ref{assumption:x-monotonicity}(i), while Panel (b) illustrates the lower bound under Assumption \ref{assumption:c-monotonicity}(i).
    The blue shaded regions indicate the possible values that $\mu_{0,\ell}(\bar x)$ can take under each assumption.
    ``L.B.'' and ``U.B.'' stand for the lower and upper bounds, respectively. The solid curves indicate the observed portions of the potential conditional mean functions. The red dot-dashed line indicates the unobserved portion of $\mu_{0,\ell}$.
    \end{flushleft}
\end{figure}

As illustrated in Figure~\ref{fig:tikz1}, under RV-monotonicity, we obtain the one-sided bound on the potential target mean $\mu_{0,\ell}(\bar{x})$ as $\mu_{0,\ell}(\bar{x})\geq \mu_{0,\ell}(\ell)$, where the right-hand side is identified as $\lim_{x\uparrow \ell}\mu_\ell(x)$ under Assumption~\ref{assumption:continuity}. 
Then, we obtain the partial identification of $\tau_{\ell}(\bar{x})$ as follows:
\begin{prop}\label{prop:rv}
    Let $\mathcal{C}=\{\ell\}$ and $\bar{x}> \ell$. Suppose that $Y_i(0)\in[M_1, M_2]$ almost surely, but the bounds are allowed to be uninformative (e.g., $M_1 = -\infty$ and $M_2 = \infty$).
    \begin{description}
        \item[(i)] Under Assumptions \ref{assumption:continuity} and \ref{assumption:x-monotonicity}(i), the sharp identified set of $\tau_{\ell}(\bar{x})$ is given by
    \begin{align}
        \left[
        \mu_{\ell}(\bar{x}) - M_2,\quad 
        \mu_{\ell}(\bar{x}) - \lim_{x\uparrow \ell}\mu_{\ell}(x)
        \right].\label{eq:bounds1}
    \end{align}

        \item[(ii)] If Assumption \ref{assumption:x-monotonicity}(ii) is assumed instead of Assumption \ref{assumption:x-monotonicity}(i), the sharp bounds are given by $[\mu_{\ell}(\bar{x}) - \lim_{x\uparrow \ell}\mu_{\ell}(x), \mu_{\ell}(\bar{x}) - M_1]$.
    \end{description}
\end{prop}


Although the upper bound in \eqref{eq:bounds1} provides nontrivial information about the treatment effect, the lower bound is not very informative in general.
To tighten the bounds, we seek an additional layer of assumptions in the subsequent sections.
We first consider identification only under monotonicity-type conditions under multi-cutoff settings, and then impose smoothness conditions for both single and multi cutoff designs.

\subsection{Multi-Cutoff RD Designs}\label{subsec:two}

This subsection studies how the identified region could be tightened in the presence of multiple cutoff points, a scenario frequently encountered in empirical work (\citealp{Cattaneo_etal:2021JASA_extrapolating}). 
For example, scholarship eligibility thresholds often vary by student background, such as gender, race, or cohort.
For expositional simplicity, our identification analysis focuses on the two-cutoff design, i.e., $\mathcal{C}=\{\ell,h\}$ with $\ell<h$. The OA extends the results to general multi-cutoff settings.

Assume that $\ell < \bar x < h$.
In the two-cutoff setting, we observe not only $\mu_\ell(x)$ but also $\mu_h(x)$, which may provide additional identification power.
In a seminal work by \cite{Cattaneo_etal:2021JASA_extrapolating}, they introduce the constant-bias assumption, which posits that $\mu_{0,h}(\bar x) - \mu_{0,\ell}(\bar x) = \mu_{0,h}(\ell) - \mu_{0,\ell}(\ell)$, and establish point identification as
\begin{align}
    \tau_\ell(\bar x) &= 
    \mu_\ell(\bar x) - \mu_{0,\ell}(\bar x) = 
    \mu_\ell(\bar x) - \mu_{0,h}(\bar x) + \left\{\mu_{0,h}(\ell) - \mu_{0,\ell}(\ell) \right\}\notag\\
    &= \mu_\ell(\bar x) - \mu_{h}(\bar x) + \mu_{h}(\ell) - \lim_{x\uparrow \ell}\mu_{\ell}(x).\label{eq:cba}
\end{align}
Although simple, this constant-bias assumption may be restrictive and empirical motivation for this assumption is unclear.

We therefore develop identification results based on alternative assumptions.
In particular, we introduce the following monotonicity condition:
\begin{assumption}[Group Monotonicity]\label{assumption:c-monotonicity}
    Either of the following conditions hold.
    \begin{description}
        \item[(i)] For $c_1,c_2 \in \mathcal C$, $c_1 \leq c_2 \Rightarrow \mu_{0,c_1}(x) \leq \mu_{0,c_2}(x)$ for $x\in[c_1, c_2]$.
        \item[(ii)] For $c_1,c_2 \in \mathcal C$, $c_1 \leq c_2 \Rightarrow \mu_{0,c_1}(x) \geq \mu_{0,c_2}(x)$ for $x\in[c_1, c_2]$.
    \end{description}
\end{assumption}
In the two-cutoff case, Assumption \ref{assumption:c-monotonicity}(i) simply states that $\mu_{0,\ell}(x) \leq \mu_{0,h}(x)$ for $x\in[\ell, h]$.
Under \textit{group monotonicity} assumption, the untreated conditional mean of the lower-cutoff group lies below that of the higher-cutoff group.
It is plausible in many multi-cutoff RD settings, especially when cutoffs are designed to reduce inequities or reflect pre-existing differences between groups.
For instance, scholarship thresholds are often relaxed for students from disadvantaged backgrounds---such as those from high-poverty regions---allowing them to qualify with lower test scores (\citealp{Melguizo_etal:2016}). Conversely, more academically prepared students may apply to competitive schools with higher cutoffs.
In both examples, the cutoff reflects differences in group characteristics, and in such cases, the group-monotonicity assumption is more likely to hold.

The next proposition summarizes identifying power of the group-monotonicity assumption alone. 
\begin{prop}\label{prop:group}
    Let $\mathcal{C}=\{\ell, h\}$ and $\bar{x} \in(\ell,h)$. Suppose that $Y_i(0)\in[M_1, M_2]$ almost surely.
    Under Assumptions \ref{assumption:continuity} and \ref{assumption:c-monotonicity}(i), the sharp identified set of $\tau_{\ell}(\bar{x})$ is given by $[\mu_{\ell}(\bar{x}) - \mu_{h}(\bar{x}),\, \mu_{\ell}(\bar{x}) - M_1]$.
    Under Assumptions \ref{assumption:continuity} and \ref{assumption:c-monotonicity}(ii), the sharp identified set is given by $[\mu_{\ell}(\bar{x}) - M_2,\, \mu_{\ell}(\bar{x}) - \mu_{h}(\bar{x})]$.
\end{prop}

The intuition behind the bounds is illustrated in Figure~\ref{fig:tikz2}.
Similarly to Proposition \ref{prop:rv}, the group-monotonicity itself is not sufficient to provide two-sided bounds.
However, when the RV-monotonicity and group-monotonicity assumptions are imposed jointly, we can obtain informative bounds on the extrapolated treatment effect, provided that the directions of monotonicity coincide.
The next theorem is our first main result:
\begin{theorem}\label{thm:main1}
    Let $\mathcal{C}=\{\ell,h\}$ and $\bar{x}\in(\ell,h)$.
    \begin{description}
        \item[(i)] Under Assumptions \ref{assumption:continuity}, \ref{assumption:x-monotonicity}(i), and \ref{assumption:c-monotonicity}(i), the sharp identified set of $\tau_\ell(\bar{x})$ is given by
        \begin{align*}
            \mathcal{B}_{\mathtt{M1}}(\bar x)
            :=
            \left[
            \mu_\ell(\bar x)
            -
            \inf_{x\in[\bar x,h)}\mu_h(x),
            \quad
            \mu_\ell(\bar x)
            -
            \lim_{x\uparrow\ell}\mu_\ell(x)
            \right].
        \end{align*}
        An analogous result under Assumptions \ref{assumption:continuity}, \ref{assumption:x-monotonicity}(ii), and \ref{assumption:c-monotonicity}(ii) is provided in Theorem A.1(i)$^\prime$ of the OA.

        \item[(ii)] Under Assumptions \ref{assumption:continuity}, \ref{assumption:x-monotonicity}(i), and \ref{assumption:c-monotonicity}(ii), the sharp identified set of $\tau_\ell(\bar{x})$ is given by
        \begin{align*}
            \mathcal{B}_{\mathtt{M2}}(\bar x)
            :=
            \left[
            \mu_\ell(\bar x)-M_2,
            \quad
            \left(
            \mu_\ell(\bar x)
            -
            \sup_{x\in[\ell,\bar x]}\mu_h(x)
            \right)
            \wedge
            \left(
            \mu_\ell(\bar x)
            -
            \lim_{x\uparrow\ell}\mu_\ell(x)
            \right)
            \right],
        \end{align*}
        provided that $Y_i(0)\in[M_1,M_2]$ almost surely.
        An analogous result under Assumptions \ref{assumption:continuity}, \ref{assumption:x-monotonicity}(ii), and \ref{assumption:c-monotonicity}(i) is provided in Theorem A.1(ii)$^\prime$ of the OA.
    \end{description}
\end{theorem}

Several important observations emerge. First, and most importantly, the identified region $\mathcal B_{\mathtt{M1}}$ is \textit{support-free} in the sense that it remains informative even without any information about the support of $Y_i(0)$ (i.e., without knowledge of $M_1$ and $M_2$).
This informativeness, however, arises only when the two monotonicity assumptions have the same direction (i.e., in Theorem~\ref{thm:main1}(i) and Theorem A.1(i)$^\prime$).
The reason for this distinction between informativeness and non-informativeness is evident from Figure~\ref{fig:mon}.
When the two monotonicity assumptions have the same direction, they provide two-sided identifying power:
For example, in case (i), RV monotonicity provides a lower bound on the unobserved $\mu_{0,\ell}(\bar x)$, whereas group monotonicity provides an upper bound on $\mu_{0,\ell}(\bar x)$.
As a result, the identified region for $\mu_{0,\ell}(\bar x)$, which is given by the intersection of the two blue shaded regions in Figure~\ref{fig:mon}, yields support-free bounds.

Second, the identified sets in Theorem~\ref{thm:main1} are in general tighter than the intersection of the bounds obtained in Propositions~\ref{prop:rv}--\ref{prop:group}. This additional identifying power arises from the interaction between the two monotonicity restrictions. For example, in case (i), RV-monotonicity and group-monotonicity imply, for every $x\in[\bar x,h)$, that $\mu_{0,\ell}(\bar x) \le \mu_{0,\ell}(x) \le \mu_{0,h}(x) = \mu_h(x)$.
It then follows that $\mu_{0,\ell}(\bar x) \le \inf_{x\in[\bar x,h)}\mu_h(x)$, rather than $\mu_{0,\ell}(\bar x) \le \mu_h(\bar x)$.

That said, the extent of this additional tightening may be limited in many empirical applications. Although Assumption~\ref{assumption:x-monotonicity} does not impose any restriction on $\mu_{0,h}(x)$ over $[\ell,h]$, the same monotonicity restriction may also plausibly hold over this region, where it is directly testable using the observed data.
If so, the infimum term in $\mathcal{B}_{\mathtt{M1}}(\bar x)$ simplifies to $\inf_{x\in[\bar x,h)}\mu_h(x) = \mu_h(\bar x)$.
Consequently, the identified sets in Theorem~\ref{thm:main1} reduce to the corresponding intersections of the bounds obtained in Propositions~\ref{prop:rv}--\ref{prop:group}, as summarized below:

\addtocounter{theorem}{-1}
\begin{theorem}[continued]
    \quad
    \begin{description}
        \item[(iii)] In each of the cases covered by Theorems \ref{thm:main1}(i)--(ii) and A.1(i)$^\prime$--(ii)$^\prime$, suppose additionally that $\mu_{0,h}$ satisfies RV monotonicity in the same direction over $[\ell,h]$. Then the sharp identified set is given by the intersection of the corresponding sharp identified sets in Propositions~\ref{prop:rv} and~\ref{prop:group}. 
        In particular, in case (i) of Theorem~\ref{thm:main1}, the sharp identified set is $\widetilde{\mathcal{B}}_{\mathtt{M1}}(\bar x) := [\mu_{\ell}(\bar{x}) - \mu_{h}(\bar{x}),\, \mu_{\ell}(\bar{x}) - \lim_{x\uparrow \ell}\mu_{\ell}(x)]$.
    \end{description}
\end{theorem}

Several additional remarks on Theorem~\ref{thm:main1}(i)--(iii) are in order.
\begin{remark}[Flexibility of the Treatment Effect Function]
    Both RV- and group-monotonicity assumptions impose restrictions only on the control state, and no assumptions are made about the functional form under the treated status. Thus, the treatment effect function itself is left unrestricted, in line with \cite{Cattaneo_etal:2021JASA_extrapolating}.
\end{remark}

\begin{remark}[Sensitivity to Transformation]
    In RD studies, it is common practice to apply a monotonic transformation to the outcome variable---for example, log transformation of annual earnings as in \cite{Oreopoulos:2006}.
    Unfortunately, the monotonicity conditions (Assumptions \ref{assumption:x-monotonicity}--\ref{assumption:c-monotonicity}) are not preserved under monotonic transformations of the outcome in general. This limitation is shared by the constant-bias assumption of \cite{Cattaneo_etal:2021JASA_extrapolating}, which is also sensitive to such transformations. Therefore, the plausibility of these assumptions should be assessed on the scale of the outcome actually used in the analysis. Section \ref{subsec:assess} discusses how to assess these assumptions.
\end{remark}

\begin{remark}[Multiple Cutoff Points]
    When more than two cutoff points are available, the identification results in Theorem~\ref{thm:main1} extend naturally, and additional comparison groups can provide further identifying power. Appendix B of the OA provides a detailed treatment. We also consider alternative group-monotonicity restrictions there.
\end{remark}

\subsection{Assessing the Plausibility of the Monotonicity Conditions}\label{subsec:assess}

The key identification assumptions (Assumptions \ref{assumption:x-monotonicity}--\ref{assumption:c-monotonicity}) are not fully testable.
That said, we can use the observed regression functions to assess the plausibility of these assumptions.
This subsection discusses such testing procedures.

\subsubsection{Testing RV and Group Monotonicity}

We focus on Assumption \ref{assumption:x-monotonicity}(i), which requires $\mu_{0,\ell}(x)$ to be weakly increasing for $x\ge \ell$; part (ii) is analogous. 
Because $\mu_{0,\ell}(x)$ is counterfactual in this region, the restriction is not directly testable. Its plausibility can nevertheless be assessed using the observed control regression function below the cutoff: violations of monotonicity of $\mu_\ell(x)$ for $x<\ell$ would cast doubt on the assumption. Similarly, in a multi-cutoff design, monotonicity of $\mu_h(x)$ for $x<h$ provides indirect evidence on the plausibility of RV monotonicity.

There are several ways to assess the monotonicity assumptions, both formally and informally.
The simplest approach is to plot the estimated regression functions not only near the cutoff points but also over a broader range of the running variable, and visually assess whether they exhibit the monotonicity patterns.

RV monotonicity can also be assessed more formally using statistical tests.
For example, one may use the test proposed by \cite{Hall_Heckman:2000} and subsequently refined and extended by \cite{Chetverikov:2019}. 
Furthermore, with an additional diffrentiability assumption, the RV-monotonicity has a testable implication that $\lim_{x\uparrow\ell}\mu_{\ell}^\prime(x)\ge0$.
The group-monotonicity condition, $\mu_{0,\ell}(x) \leq \mu_{0,h}(x)$ for $x\in[\ell,h]$, yields the testable implication that $\lim_{x\uparrow\ell}\mu_{\ell}(x) \leq \mu_{h}(\ell)$, which can be assessed using standard nonparametric estimators.
In addition, the assumption can be evaluated indirectly. For example, the inequality $\mu_{\ell}(x) \leq \mu_h(x)$ over $(-\infty,\ell)$ can be examined both graphically and formally.

\subsection{Constant-Bias Extrapolation as a Benchmark}

The preceding plausibility checks also provide a useful connection between our bounds and the point-identification approach of \cite{Cattaneo_etal:2021JASA_extrapolating}.
In particular, when the observed monotonicity patterns are aligned with the monotonicity restrictions imposed in Theorem \ref{thm:main1}, the point-identified value under the constant-bias assumption is contained in our identified set.

To see this, recall \eqref{eq:cba} and let $\mu_{0,\ell}^{\mathtt{CB}}(\bar x) := \mu_h(\bar x) - \mu_h(\ell) + \lim_{x\uparrow\ell}\mu_\ell(x)$ denote the counterfactual untreated mean implied by the constant-bias assumption.
Notice that
\begin{align*}
    \mu_{0,\ell}^{\mathtt{CB}}(\bar x) - \mu_h(\bar x) 
    = \lim_{x\uparrow\ell}\mu_\ell(x)-\mu_h(\ell),
    \quad \text{and}\quad
    \mu_{0,\ell}^{\mathtt{CB}}(\bar x) - \lim_{x\uparrow\ell}\mu_\ell(x)
    = \mu_h(\bar x)-\mu_h(\ell).
\end{align*}
The first equality shows that if group monotonicity holds at the cutoff, then the constant-bias-implied counterfactual $\mu_{0,\ell}^{\mathtt{CB}}$ satisfies the same group ordering throughout $[\ell,h)$. 
The second equality shows that if $\mu_h$ satisfies RV monotonicity over $[\ell,h)$ in the same direction as imposed on the lower-cutoff group, then $\mu_{0,\ell}^{\mathtt{CB}}$ inherits that RV-monotonicity direction. Hence, under these observable monotonicity patterns, $\mu_{0,\ell}^{\mathtt{CB}}$ satisfies both monotonicity restrictions.

More concretely, suppose the directions considered in Theorem \ref{thm:main1}(i) and further assume that $\mu_{0,h}$ satisfies the RV monotonicity over the region $[\ell, h)$.
Then $\lim_{x\uparrow\ell}\mu_\ell(x) \le \mu_h(\ell) \le \mu_h(\bar x)$, where the first inequality uses group monotonicity. This in turn implies $\lim_{x\uparrow\ell}\mu_\ell(x) \le \mu_{0,\ell}^{\mathtt{CB}}(\bar x) \le \mu_h(\bar x)$.
Therefore, the point-identified treatment effect under the constant-bias assumption satisfies,
\begin{align*}
    \tau_\ell^{\mathtt{CB}}(\bar x)
    :=
    \mu_\ell(\bar x)-\mu_{0,\ell}^{\mathtt{CB}}(\bar x)
    \in \bigg[
            \mu_{\ell}(\bar{x}) - \mu_{h}(\bar{x}),\quad
            \mu_{\ell}(\bar{x}) - \lim_{x\uparrow \ell}\mu_{\ell}(x)
            \bigg] = \widetilde{\mathcal{B}}_{\mathtt{M1}}(\bar x) = {\mathcal{B}}_{\mathtt{M1}}(\bar x).
\end{align*}
An analogous result holds for the monotonicity directions in Theorem A.1(i)$^\prime$.
For part (ii), the same argument shows that the constant-bias-implied value satisfies the monotonicity restrictions.
It therefore belongs to the identified set whenever it also respects the relevant support restriction, $\mu_{0,\ell}^{\mathtt{CB}}(\bar x)\leq M_2$.

These observations provide a natural sensitivity-analysis interpretation of our bounds. When the observed group ordering and the RV-monotonicity pattern of $\mu_h$ are consistent with the corresponding monotonicity restrictions imposed on the counterfactual regression function, $\tau_\ell^{\mathtt{CB}}(\bar x)$ can be viewed as a particular extrapolation lying within the broader range of counterfactual values permitted by our monotonicity restrictions. Our bounds can therefore be used together with the point-identified value under the constant-bias assumption to conduct a sensitivity analysis of how conclusions change when the constant-bias restriction is relaxed while preserving the corresponding monotonicity patterns.

\section{Tightening Bounds by Smoothness Assumptions}\label{sec:smooth}

The bounds in Theorem \ref{thm:main1}, obtained under the two monotonicity assumptions, will be credible in many applications. A potential drawback, however, is that they may still be too wide to be informative. 
To understand what drives their width, consider the particularly relevant case in which $\mu_h$ is monotone and recall the bounds $\widetilde{\mathcal B}_{\mathtt{M1}}(\bar x)$ in Theorem~\ref{thm:main1}(iii). Observe that
\begin{align*}
    |\widetilde{\mathcal B}_{\mathtt{M1}}(\bar x)| = \mu_{0,h}(\bar{x}) - \mu_{0,\ell}(\ell)
    = \underbrace{\left\{\mu_{0,h}(\bar{x}) - \mu_{0,h}(\ell)\right\}}_{\text{growth from $\ell$ for group $h$}}
    + \underbrace{\left\{\mu_{0,h}(\ell) - \mu_{0,\ell}(\ell)\right\}}_{\text{group difference at $\ell$}}.
\end{align*}
Thus, the bounds tend to be less informative when $\mu_{0,h}$ increases rapidly between $\ell$ and $\bar{x}$, or when the cross-group difference $\mu_{0,h}(\ell) - \mu_{0,\ell}(\ell)$ is large.

One potentially undesirable feature of $\widetilde{\mathcal B}_{\mathtt{M1}}(\bar x)$ is that the identified set may expand abruptly from point identification immediately to the right of the cutoff $\ell$. More precisely, for $\bar{x}=\ell+\varepsilon$ with a small $\varepsilon>0$,
\begin{align*}
    |\widetilde{\mathcal B}_{\mathtt{M1}}(\ell + \varepsilon)| 
    = \underbrace{\left\{\mu_{0,h}(\ell + \varepsilon) - \mu_{0,h}(\ell)\right\}}_{\approx 0 \text{ if } \varepsilon \approx 0}
    + \left\{\mu_{0,h}(\ell) - \mu_{0,\ell}(\ell)\right\}.
\end{align*}
Provided that $\mu_{0,h}$ is continuous at $\ell$, the first term converges to zero as $\varepsilon\downarrow 0$, whereas the second term remains fixed. Consequently, the width of the identified set does not converge to zero but instead converges to the cross-group difference $\mu_{0,h}(\ell)-\mu_{0,\ell}(\ell)$.
The bounds therefore permit $\mu_{0,\ell}$ to change sharply immediately to the right of $\ell$, even though such behavior may be implausible in many empirical applications.

This concern can be formalized by imposing a smoothness restriction on $\mu_{0,\ell}$. Such a restriction limits abrupt changes in the unobserved conditional mean function and can therefore tighten the identified set. This section studies the identifying power of smoothness restrictions when combined with the monotonicity conditions.

\subsection{Smoothness Restriction on the Conditional Mean Function}

In the recent RD and nonparametrics literature, it is common to assume that the conditional mean function $\mu_{0,\ell}$ belongs to a class of smooth functions.
Motivated by its intuitive appeal and suitability for our extrapolation setting, this subsection focuses on the Lipschitz class considered by \cite{Kwon_Kwon:2020}:
\begin{align*}
    \mathcal{L}(C) = \left\{
    \mu : \left|\mu(x_1) - \mu(x_2)\right| \leq C|x_1 - x_2| \text{ for all } x_1, x_2 \in [\ell, \infty) \cap \mathcal X
    \right\}.
\end{align*}
When $\mu$ is differentiable, $\mu \in \mathcal{L}(C)$ simply says that $|\mu^\prime| \le C$. We now introduce the following smoothness assumption on $\mu_{0,\ell}$:
\begin{assumption}[Bounded Slope]\label{assumption:lipchitz-l}
    It holds that $\mu_{0,\ell} \in \mathcal{L}(C_\ell)$ for a known constant $C_\ell>0$.
\end{assumption}
Notice that we do not impose a corresponding smoothness restriction on $\mu_{1,\ell}$, so as not to restrict the treatment-effect function.

Assumption \ref{assumption:lipchitz-l} provides an additional constraint on the counterfactual conditional mean, potentially tightening the bounds established in the previous section.
Because it does not rely on the existence of multiple cutoffs, Assumption \ref{assumption:lipchitz-l} is particularly useful in settings with a single cutoff.
The following proposition presents the identification result for the single-cutoff case.
\begin{prop}\label{prop:smooth-single}
    Let $\mathcal{C}=\{\ell\}$ and $\bar{x}> \ell$.
    Suppose that $Y_i(0)\in[M_1, M_2]$ almost surely.
    \begin{enumerate}
        \item[(i)] Under Assumptions \ref{assumption:continuity} and \ref{assumption:lipchitz-l}, the sharp identified set of $\tau_{\ell}(\bar{x})$ is given by
        \begin{align*}
            \mathcal{B}_{\mathtt{BS}}(\bar x) := \bigg[
            &\mu_{\ell}(\bar{x}) - \left(M_2\,\,\wedge \,\,
                \left\{\lim_{x\uparrow \ell}\mu_{\ell}(x) + C_\ell(\bar{x}-\ell)\right\}\right),\\
            &\quad \mu_{\ell}(\bar{x}) - \left(M_1\,\,\vee \,\,
                \left\{\lim_{x\uparrow \ell}\mu_{\ell}(x) - C_\ell(\bar{x}-\ell)\right\}\right)
            \bigg].
        \end{align*}

        \item[(ii)] Under Assumptions \ref{assumption:continuity}, \ref{assumption:x-monotonicity}(i), and \ref{assumption:lipchitz-l}, the sharp identified set of $\tau_{\ell}(\bar{x})$ is given by the intersection of $\mathcal{B}_{\mathtt{BS}}(\bar x)$ and the sharp identified set in Proposition~\ref{prop:rv}(i).
        Under Assumptions \ref{assumption:continuity}, \ref{assumption:x-monotonicity}(ii), and \ref{assumption:lipchitz-l}, the sharp identified set of $\tau_{\ell}(\bar{x})$ is given by the intersection of $\mathcal{B}_{\mathtt{BS}}(\bar x)$ and the sharp identified set in Proposition~\ref{prop:rv}(ii).
    \end{enumerate}
\end{prop}

\subsection{Smoothness Restriction on the Bias Function}\label{subsec:smoothbias}

In the presence of multiple cutoffs, a natural restriction is to impose a smoothness restriction on what \cite{Cattaneo_etal:2021JASA_extrapolating} calls ``bias'' $\mu_{0,c}(x) - \mu_{0,\ell}(x) =: \delta_c(x)$ for $c>\ell$.
We employ the following smooth-bias assumption proposed in \citet[Assumption 2]{zhang:2024}, a relaxation of the constant-bias assumption:
\begin{assumption}[Smooth Bias]\label{assumption:lipchitz-bias}
    For $c\in\mathcal C\backslash\{\ell\}$, $\delta_c \in \mathcal{L}(B_c)$, where $B_c\geq 0$ is a known constant.
\end{assumption}
For $c=h$, Assumption \ref{assumption:lipchitz-bias} posits that $\left|\delta_h(\bar{x}) - \delta_h(\ell)\right| \leq B_h(\bar{x} - \ell)$. The constant-bias assumption is a special case with $B_h=0$.
The next theorem extends Theorem~\ref{thm:main1} by additionally imposing the smooth-bias assumption:
\begin{theorem}\label{thm:main2}
Let $\mathcal{C}=\{\ell,h\}$ and $\bar{x}\in(\ell,h)$.
\begin{description}
    \item[(i)] Under Assumptions \ref{assumption:continuity} and \ref{assumption:lipchitz-bias}, the sharp identified set of $\tau_{\ell}(\bar{x})$ is given by
    \begin{align*}
        \mathcal B_{\mathtt S}(\bar x) := \left[
        \tau_\ell^{\mathtt{CB}}(\bar x)-B_h(\bar x-\ell),\,
        \tau_\ell^{\mathtt{CB}}(\bar x)+B_h(\bar x-\ell)
        \right],
    \end{align*}
    where $\tau_\ell^{\mathtt{CB}}(\bar x) = \mu_\ell(\bar x)-\mu_h(\bar x)+\mu_h(\ell)-\lim_{x\uparrow\ell}\mu_\ell(x)$.

    \item[(ii)] Under Assumptions \ref{assumption:continuity}, \ref{assumption:x-monotonicity}(i), \ref{assumption:c-monotonicity}(i), and \ref{assumption:lipchitz-bias}, the sharp identified set of $\tau_{\ell}(\bar{x})$ is given by
    \begin{align*}
        \mathcal B_{\mathtt{MS1}}(\bar x)
        :=
        &\bigg[
        \left\{\tau_\ell^{\mathtt{CB}}(\bar x) - B_h(\bar x-\ell)\right\}
        \vee
        \sup_{x\in[\bar x,h)} \left\{\mu_\ell(\bar x) - \mu_h(x) + V_{B_h}^{+}(\bar x,x)\right\},\\
        &\quad
        \mu_\ell(\bar x) - \lim_{x\uparrow\ell}\mu_\ell(x) - V_{B_h}^{+}(\ell,\bar x)
        \bigg],
    \end{align*}
    where
    \begin{align*}
        V_{B_h}^{+}(s,t)
        :=
        \sup_{\substack{K\geq1,\\ s=x_0 \le x_1 \le \cdots \le x_K=t}}
        \sum_{k=1}^K \left[\mu_h(x_k)-\mu_h(x_{k-1}) - B_h(x_k-x_{k-1})\right]_+.
    \end{align*}
    An analogous result under Assumptions \ref{assumption:continuity}, \ref{assumption:x-monotonicity}(ii), and \ref{assumption:c-monotonicity}(ii) is provided in Theorem A.2(ii)$^\prime$ of the OA.

    \item[(iii)] Under Assumptions \ref{assumption:continuity}, \ref{assumption:x-monotonicity}(i), \ref{assumption:c-monotonicity}(ii), and \ref{assumption:lipchitz-bias}, the sharp identified set of $\tau_{\ell}(\bar{x})$ is given by
    \begin{align*}
        \mathcal B_{\mathtt{MS2}}(\bar x)
        :=
        &\bigg[
        \left\{\tau_\ell^{\mathtt{CB}}(\bar x) -B_h(\bar x-\ell)\right\}
        \vee
        \sup_{x\in[\bar x,h)} \left\{\mu_\ell(\bar x) - M_2 + V_{B_h}^{+}(\bar x,x)
        \right\},\\
        &\quad
        \left\{\mu_\ell(\bar x) - \lim_{x\uparrow\ell}\mu_\ell(x) - V_{B_h}^{+}(\ell,\bar x)\right\}
        \wedge
        \inf_{x\in[\ell,\bar x]} \left\{\mu_\ell(\bar x) - \mu_h(x) - V_{B_h}^{+}(x,\bar x)\right\}
        \bigg],
    \end{align*}
    provided that $Y_i(0)\in[M_1,M_2]$ almost surely.
    An analogous result under Assumptions \ref{assumption:continuity}, \ref{assumption:x-monotonicity}(ii), and \ref{assumption:c-monotonicity}(i) is provided in Theorem A.2(iii)$^\prime$ of the OA.
\end{description}
\end{theorem}

\begin{remark}
    The identified sets in Theorem~\ref{thm:main2} converge to the point-identified constant-bias specification of \cite{Cattaneo_etal:2021JASA_extrapolating} as the smoothness parameter $B_h \to 0$ whenever the constant-bias-implied counterfactual $\mu_{0,\ell}^{\mathtt{CB}}$ satisfies the remaining monotonicity (and support) restrictions. 
    We therefore propose the bounds in Theorem~\ref{thm:main2} as our baseline specification, treating $B_h$ as a sensitivity parameter.
\end{remark}
\begin{remark}
    Theorem~\ref{thm:main2}(i) implies that $|\mathcal B_{\mathtt{S}}(\bar x)| = 2B_h(\bar x-\ell)$. Hence, the bounds under the smooth-bias assumption are particularly informative near $\ell$, thereby mitigating a limitation of the bounds obtained under the monotonicity restrictions. However, for a fixed $B_h$, the width of the smooth-bias bounds increases with the extrapolation distance.
\end{remark}

Under the assumptions made in Theorem \ref{thm:main2}(ii), $\mathcal B_{\mathtt{M1}}(\bar x) \cap \mathcal B_{\mathtt S}(\bar x)$ and $\widetilde{\mathcal B}_{\mathtt{M1}}(\bar x) \cap \mathcal B_{\mathtt S}(\bar x)$ are valid identified sets of $\tau_{\ell}(\bar{x})$.
The same is true for case (iii) of Theorem \ref{thm:main2}.
However, these intersection bounds are not sharp in general.

The additional tightness of $\mathcal B_{\mathtt{MS1}}$ and $\mathcal B_{\mathtt{MS2}}$ is reflected by the presence of $V_{B_h}^+$.
To see why this term arises, let us focus on $\mathcal B_{\mathtt{MS1}}$.
Note that, for any $\ell\leq s \le t<h$, the smooth-bias restriction implies $\mu_{0,\ell}(t)-\mu_{0,\ell}(s) = \mu_h(t)-\mu_h(s) - \{\delta_h(t)-\delta_h(s)\} \geq \mu_h(t) - \mu_h(s) - B_h(t-s)$, while RV-monotonicity implies $\mu_{0,\ell}(t)-\mu_{0,\ell}(s)\geq0$.
Hence, $\mu_{0,\ell}(t)-\mu_{0,\ell}(s) \ge \left[\mu_h(t) - \mu_h(s) - B_h(t-s)\right]_+$.
Applying this inequality along any partition $s=x_0 \le x_1 \le \cdots \le x_K=t$ and summing gives $\mu_{0,\ell}(t)-\mu_{0,\ell}(s) \ge \sum_{k=1}^K [\mu_h(x_k) - \mu_h(x_{k-1}) - B_h(x_k-x_{k-1})]_+$.
We can set $s=\ell$ and $t=\bar x$.
Then taking the supremum over all such partitions yields, together with continuity, $\mu_{0,\ell}(\bar x) \geq \lim_{x\uparrow\ell}\mu_\ell(x) + V_{B_h}^{+}(\ell,\bar x)$.
Similarly, $\mu_{0,\ell}(t)-\mu_{0,\ell}(s)\geq V_{B_h}^{+}(s,t)$ and group-monotonicity imply $\mu_h(x) \geq \mu_{0,\ell}(x) \ge \mu_{0,\ell}(\bar x)+V_{B_h}^{+}(\bar x,x)$ for every $x\in[\bar x,h)$.
Then we have $\mu_{0,\ell}(\bar x) \leq \mu_h(x)-V_{B_h}^{+}(\bar x,x)$ for every $x\in[\bar x,h)$.
These arguments show the validity of the tighter bounds, and Theorem~\ref{thm:main2} confirms the bounds are sharp as well.

If $\mu_{0,h}$ is absolutely continuous, we can show that the intersection bounds $\mathcal B_{\mathtt{M1}}(\bar x) \cap \mathcal B_{\mathtt S}(\bar x)$ and the sharp bounds $\mathcal B_{\mathtt{MS1}}(\bar x)$ coincide if the sign of $\mu_h^\prime(x) - B_h$ does not change sign almost everywhere on $[\ell,h)$. 
We also note that the lower endpoints of $\mathcal B_{\mathtt{M1}}(\bar x) \cap \mathcal B_{\mathtt S}(\bar x)$ and $\mathcal B_{\mathtt{MS1}}(\bar x)$ coincide if $\mu_{h}$ is weakly increasing over $[\bar x, h)$.
See Section A.2 of the OA for details.

\subsubsection{Numerical Illustration}

\begin{figure}[t]
    \centering
    \includegraphics[width=0.75\linewidth]{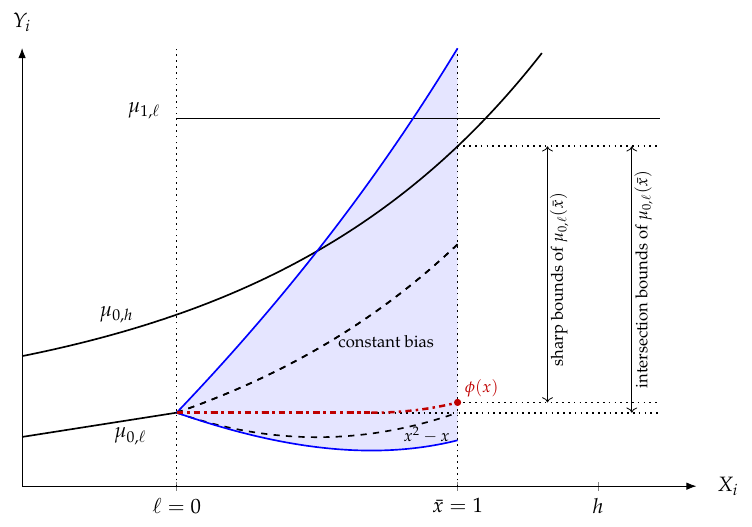}
    \caption{Numerical Illustration: Sharp vs. Intersection Bounds}
    \label{fig:sharp}

    \begin{flushleft}
    \footnotesize
    \renewcommand{\baselineskip}{11pt}
    \textbf{Note:} $\mu_{0,h}(x)=\exp(x)$, $\mu_{0,\ell}(\ell)=0$, and $\mu_{1,\ell}(x) = 3$.
    The blue shaded area indicates the range of values that $\mu_{0,\ell}$ can take under the smooth-bias assumption.
    The red dot-dashed curve $\phi$ is given by $\phi(x):=\lim_{z\uparrow\ell}\mu_{\ell}(z) + V_{B_h}^{+}(\ell,x)$, where, for $B_h=2$, $V_{B_h}^{+}(\ell,x)$ is calculated as $V_{B_h}^{+}(\ell,x) = \int_{\log(2)}^x (\exp(z) - B_h)\,dz = \exp(x)-B_h x -2 + B_h\log(2)$ for $x\ge \log(2)\approx 0.69$ and $V_{B_h}^{+}(\ell,x) = 0$ for $x<\log(2)$.
    \end{flushleft}
\end{figure}

A numerical example may clearly illustrate why $\mathcal B_{\mathtt{M1}}(\bar x) \cap \mathcal B_{\mathtt S}(\bar x)$ need not be sharp.
Suppose that $\ell=0$ and $\bar x=1$, and let $\mu_{0,h}(x)=\exp(x)$ on $[0,h]$ for some $h>1$.
Also suppose that $\mu_{0,\ell}(\ell)=\mu_{0,\ell}(0)=0$. We specify $B_h = 2$.
See Figure~\ref{fig:sharp} for the corresponding graphical illustration.

We are interested in the identified set for $\mu_{0,\ell}(\bar x)$.
To illustrate the interaction between the two restrictions, let us consider whether $\mu_{0,\ell}(\bar x)=0$ is attainable.
Note that, under the RV-monotonicity assumption alone, $\mu_{0,\ell}(\bar x)=0$ is admissible (only) if $\mu_{0,\ell}(x)=0$ for all $x\in[0,1]$.
In contrast, under the smooth-bias assumption alone with $B_h=2$, $\mu_{0,\ell}(\bar x)=0$ is admissible; for example, it is attained by $\mu_{0,\ell}(x)=x^2-x$ for $x\in[0,1]$.

Now suppose that the two assumptions are imposed jointly.
Note that $\mu_{0,h}^\prime(1)=\exp(1)\approx 2.7>B_h$.
Hence, under the smooth-bias restriction, which requires $\mu_{0,h}^\prime(1) - \mu_{0,\ell}^\prime(1) \le B_h$, $\mu_{0,\ell}^\prime(1)$ must be strictly positive.
On the other hand, if $\mu_{0,\ell}(1)=0$, RV-monotonicity requires $\mu_{0,\ell}(x)=0$ for all $x\in[0,1]$, which implies $\mu_{0,\ell}^\prime(1)=0$.
Thus, $\mu_{0,\ell}(1)=0$ cannot satisfy the two restrictions simultaneously.

In words, for $\mu_{0,\ell}(1)=0$ to be compatible with the smooth-bias restriction, $\mu_{0,\ell}$ must fall below zero somewhere on $[0,1]$ before increasing sufficiently rapidly near $x=1$.
RV monotonicity, however, rules out such negative values because $\mu_{0,\ell}(0)=0$.
Therefore, $\mu_{0,\ell}(1)=0$ is ruled out under the joint restrictions, even though it is admissible under either restriction imposed separately.

\subsection{Incorporating More General Smoothness Restrictions}\label{subsec:general}

We have covered several leading cases in the previous subsections, but in some applications researchers may wish to impose alternative smoothness restrictions or combinations of different restrictions. These additional restrictions can be incorporated straightforwardly into the framework developed above, although the sharpness of the resulting identified set is not guaranteed in general. In particular, suppose that there are $K$ sets of identifying assumptions, each of which yields a valid identified set $\tau_\ell(\bar x) \in \mathcal{B}_k$ for $k=1,\ldots,K$. Under the joint imposition of these assumptions, $\bigcap_{k=1}^K \mathcal{B}_k$ provides a weakly tighter valid identified set.

For example, suppose a canonical single-cutoff design and assume that $\mu_{0,\ell}^\prime$ is continuous at $\ell$ and that $\mu_{0,\ell}\in\mathcal H(C_\ell)$ for a known constant $C_\ell>0$, where
\begin{align*}
    \mathcal H(C)
    :=
    \left\{
    \mu:
    |\mu^\prime(x_1)-\mu^\prime(x_2)|
    \leq C|x_1-x_2|
    \text{ for all }x_1,x_2\in[\ell,\infty)\cap\mathcal X
    \right\}.
\end{align*}
This type of smoothness condition is considered in \cite{Kolear_Rothe:2018}; \cite{Rambachan_Roth:2023} considers a similar restriction in the DID context. The sharp identified set under this smoothness restriction alone is then $\mathcal B_H(\bar x) := [\mu_\ell(\bar x) - \lim_{x\uparrow\ell}\mu_\ell(x) - \lim_{x\uparrow\ell}\mu_\ell^\prime(x)(\bar x-\ell) \pm {C_\ell}(\bar x-\ell)^2/{2}]$.
If RV monotonicity (Assumption~\ref{assumption:x-monotonicity}) is also imposed, intersecting $\mathcal B_H(\bar x)$ with the identified set implied by RV monotonicity in \eqref{eq:bounds1} yields a valid identified set.

As another example, suppose that Assumptions \ref{assumption:continuity}, \ref{assumption:x-monotonicity}(i), \ref{assumption:c-monotonicity}(i), \ref{assumption:lipchitz-l}, and \ref{assumption:lipchitz-bias} hold.
Then $\mathcal B_{\mathtt{M}}(\bar x) \cap \mathcal B_{\mathtt{BS}}(\bar x) \cap \mathcal{B}_{\mathtt{S}}(\bar x)$ is a valid identified set for $\tau_\ell(\bar x)$.

Thus, when obtaining tighter valid bounds is the primary objective and sharpness is of secondary importance, researchers can construct a weakly smaller valid identified set by intersecting the baseline identified set with those obtained under additional restrictions.
In general, the intersection bounds need not be sharp, for the same reason discussed in Section~\ref{subsec:smoothbias}. Obtaining sharp bounds would require formally characterizing the additional identifying content generated by the interaction among the imposed assumptions, which generally needs to be done on a case-by-case basis. Nevertheless, from a practical perspective, the intersection bounds can still be useful for tightening the baseline identified sets.
Furthermore, this construction has the practical advantage that standard moment-inequality-based inference can often be applied directly, as discussed in the following section.

\section{Implementation}\label{sec:inf}


\subsection{Estimation}

The identified sets provided above depend on the values of regression functions at several evaluation points. 
These values can be estimated using standard nonparametric regression methods. 
Alternatively, given our focus on identification under monotonicity restrictions on the regression functions, shape-constrained methods such as isotonic regression or rearrangement-based estimators can also be employed (e.g., \citealp{Babii_Kumar:2023, Chernozhukov09}).

\subsection{Bias-Aware Confidence Intervals}

A natural approach to statistical inference in our setting is the bias-aware (or honest) inference framework. In particular, \citet{Armstrong_Kolesar:2018} develop a general theory for inference on linear functionals of regression functions that are known to belong to a convex function class. Our setting is well suited to this framework because our target parameter is written by linear functionals of regression functions, and our main restrictions---monotonicity, Lipschitz-type smoothness, and combinations thereof---define convex classes of regression functions. Accordingly, the general procedure developed in \citet[Section~3]{Armstrong_Kolesar:2018} can be applied to inference on the extrapolated treatment effect.

An important advantage of this approach is that it allows us to retain the identifying content of the joint restrictions without requiring an explicit characterization of the corresponding sharp identified set. More precisely, the modulus-of-continuity problem is solved directly over the function class satisfying all of the maintained restrictions simultaneously. It therefore automatically incorporates interactions among the identifying restrictions, including those that give rise to terms such as $V_{B_h}^{+}$ in Theorem~\ref{thm:main2}. 

The corresponding practical consideration is that the modulus-of-continuity problem must generally be solved for each particular function class on a case-by-case basis. For general combinations of restrictions, the characterization of the least-favorable functions and the associated modulus can be complex. 
In such a case, solving the modulus problem over a convex outer relaxation may be computationally useful (see Remark~\ref{remark:poly} below).

\subsection{Inference Using Moment Inequalities}

An alternative approach is to conduct inference directly on an outer approximation to the identified set using moment inequalities. Even when a sharp characterization of the identified set is analytically complicated, a valid outer identified set often admits a simpler representation. 
We focus on the cases where the (potentially outer) identified set can be represented as $\Theta(\beta):=\{\theta\in\mathbb R:r_\theta\leq G \beta\}$, where $r_\theta$ and $G$ are a known vector and matrix, respectively, and $\beta\in\mathbb R^p$ is a finite-dimensional vector of identified features of the observed regression functions.
Some examples are in order:
\begin{example}\label{ex:main}
    Suppose the setup in Theorem~\ref{thm:main2}(ii), that is, Assumptions \ref{assumption:continuity}, \ref{assumption:x-monotonicity}(i), \ref{assumption:c-monotonicity}(i), and \ref{assumption:lipchitz-bias} hold true. In this case, $\widetilde{\mathcal B}_{\mathtt{M1}}(\bar x)\cap\mathcal B_{\mathtt S}(\bar x)$ is a valid (outer) identified set.
    Define the vector of identified quantities $\beta = (\mu_\ell(\bar x), \lim_{x\uparrow\ell}\mu_\ell(x), \mu_h(\bar x), \mu_h(\ell))^\top$.
    Then $\widetilde{\mathcal B}_{\mathtt{M1}}(\bar x)\cap\mathcal B_{\mathtt S}(\bar x)$ admits the representation $\Theta(\beta) = \{\theta \in \mathbb{R} : a\theta - b \le G\beta\}$, where
    \begin{align*}
        a
        =
        \begin{pmatrix}
            -1\\
             1\\
            -1\\
             1
        \end{pmatrix},\quad
        b
        =
        \begin{pmatrix}
            0\\
            0\\
            B_h (\bar x - \ell)\\
            B_h (\bar x - \ell)
        \end{pmatrix},\quad
        G
        =
        \begin{pmatrix}
            -1& 0&  1& 0\\
             1&-1&  0& 0\\
            -1& 1&  1&-1\\
             1&-1& -1& 1
        \end{pmatrix}.
    \end{align*}
\end{example}

\begin{example}\label{ex:outer-approx}
    Suppose the setup in Theorem~\ref{thm:main1}(i), that is, Assumptions \ref{assumption:continuity}, \ref{assumption:x-monotonicity}(i), and \ref{assumption:c-monotonicity}(i) hold true. The sharp identified set $\mathcal{B}_{\mathtt{M1}}(\bar x) = [\mu_\ell(\bar x) - \inf_{x\in[\bar x,h)}\mu_h(x), \, \mu_\ell(\bar x) - \lim_{x\uparrow\ell}\mu_\ell(x)]$ satisfies 
    \begin{align*}
        \mathcal{B}_{\mathtt{M1}}(\bar x) \subseteq 
        \overline{\mathcal{B}}_{\mathtt{M1}}(\bar x) :=
        \left[
            \mu_\ell(\bar x) - \min_{x\in \mathcal D}\mu_h(x),
            \quad\mu_\ell(\bar x) - \lim_{x\uparrow\ell}\mu_\ell(x)
        \right],
    \end{align*}
    where $\mathcal D := \{x_1,\ldots,x_D\} \subset [\bar x,h)$ is pre-specified finite grid points.
    Define the vector of identified quantities $\beta = (\mu_\ell(\bar x), \lim_{x\uparrow\ell}\mu_\ell(x), \mu_h(x_1),\ldots,\mu_h(x_D))^\top$.
    Then $\overline{\mathcal{B}}_{\mathtt{M1}}(\bar x)$ admits the representation $\Theta(\beta) = \{\theta \in \mathbb{R} : a\theta - b \le G\beta\}$, where
    \begin{align*}
        a
        =
        \begin{pmatrix}
            -1\\
            \vdots\\
            -1\\
             1
        \end{pmatrix},\quad
        b
        =
        \begin{pmatrix}
            0\\
            \vdots\\
            0\\
            0
        \end{pmatrix},\quad
        G
        =
        \begin{pmatrix}
            -1 & 0 & 1 & 0 & \cdots & 0\\
            -1 & 0 & 0 & 1 & \cdots & 0\\
            \vdots & \vdots & \vdots & \vdots & \ddots & \vdots\\
            -1 & 0 & 0 & 0 & \cdots & 1\\
             1 &-1 & 0 & 0 & \cdots & 0
        \end{pmatrix}.
    \end{align*}
\end{example}

This representation naturally leads to a moment-inequality-based inference procedure.
To illustrate, suppose that $\widehat\beta \sim \mathcal N(\beta,\Sigma_\beta)$.
For any candidate value $\theta$, it follows that $r_\theta-G\widehat\beta \sim \mathcal N(r_\theta-G\beta,\,G\Sigma_\beta G^\top)$.
Hence, the null hypothesis that $\theta\in\Theta(\beta)$ is equivalent to the Gaussian mean restriction $\mathbb{E}[r_\theta-G\widehat\beta]\leq 0$.
Thus, for each candidate value of $\theta$, inference reduces to testing a finite-dimensional system of inequalities on the mean of a Gaussian random vector.
This Gaussian mean-inequality formulation provides an alternative route to valid inference that avoids solving the modulus-of-continuity problem.

\citet{Andrews_etal:2023} provides inference procedures that can be applied to this problem and establishes asymptotic validity without requiring exact Gaussianity.
\citet[Section~3.3]{Rambachan_Roth:2023} establishes asymptotic validity in a DID context for a special case of the general framework in \citet{Andrews_etal:2023} that nevertheless nests the inference setup considered here.
Accordingly, the same inference procedure applies to our setting once the corresponding regularity conditions are satisfied.
In our setting, the required uniform Gaussian approximation for the local-polynomial estimators follows under standard conditions from existing robust-bias-corrected local-polynomial theory (\citealp{calonico2022coverage}; see also \citealp{Calonico_etal:2014, calonico2018effect}).

In our empirical application below, we employ this moment-inequality-based inference procedure.

\begin{remark}
    We note that the asymptotic normality condition on $\widehat\beta$ excludes some useful classes of estimators. For example, the isotonic regression estimator of \cite{Babii_Kumar:2023} is known to have a nonstandard asymptotic distribution, and hence the inference framework outlined here does not apply directly. 
\end{remark}
\begin{remark}\label{remark:poly}
    The representation $\Theta(\beta)$ can be useful for the bias-aware framework as well.
    One may treat the restrictions defining $\Theta(\beta)$ as a convex outer relaxation of the original restrictions and solve the corresponding modulus-of-continuity problem over the relaxed class. 
    Relative to a modulus problem based on the original class, this approach generally discards some identifying information, but it may simplify the characterization and computation of the modulus.
\end{remark}

\section{Empirical Illustration}\label{sec:emp}

We revisit the empirical analysis of \cite{Cattaneo_etal:2021JASA_extrapolating}.
They investigated the extrapolated effect of the \textit{Acceso con Calidad a la Educación Superior} (ACCES) program---a national merit-based financial aid initiative---on higher education enrollment among Colombian students, originally studied by \cite{Melguizo_etal:2016}.
Eligibility for ACCES requires students to score above a specific threshold on the national high school exit exam (SABER 11). The score ranges from 1 (best) to 1000 (worst), and the eligibility cutoff was fixed at 850 prior to 2008. Beginning in 2009, however, region-specific cutoffs were introduced, resulting in a multi-cutoff RD setting. For consistency, we multiply the scores by $-1$, so that higher achievement corresponds to larger values.
\cite{Cattaneo_etal:2021JASA_extrapolating} leveraged this design to estimate extrapolated RD effects on college enrollment, focusing on two cohorts: students who applied between 2000 and 2008 (cutoff $-850$) and those who applied between 2009 and 2010 in a region with a cutoff of $-571$. 

\subsection{Identifying Assumptions}

The subsequent analysis reports the identified set under Assumptions \ref{assumption:continuity}, \ref{assumption:x-monotonicity}(i), \ref{assumption:c-monotonicity}(i), and \ref{assumption:lipchitz-bias}.
We first discuss and assess the plausibility of these assumptions.

\subsubsection{Monotonicity Assumptions}

The RV-monotonicity assumption is plausible in this setting. The assumption that $\mu_{0,\ell}$ is monotonically increasing implies that students with higher SABER 11 scores are more likely to enroll in college, which is a natural restriction in this context.

The group-monotonicity assumption also appears reasonable in the ACCES setting. A key institutional feature is that the 2009 reform introduced cutoffs progressively: regions with greater socioeconomic disadvantage were assigned lower (i.e., lenient) thresholds, whereas more advantaged regions faced higher (i.e., stringent) thresholds. Notably, the region with the higher cutoff of $h=-571$ is among the most advantaged, with a very small share of students from low-socioeconomic-status backgrounds \citep[Figure 1]{Melguizo_etal:2016}. It is plausible that students from higher-SES backgrounds have, on average, a higher probability of enrolling in higher education even at the same SABER 11 scores, for example because of better access to universities and other educational institutions, greater financial support from their families, or greater awareness of the returns to human-capital investment. If so, the group-monotonicity restriction is plausible.

\subsubsection{Constant- and Smooth-Bias Assumptions}

In contrast, the plausibility of the exact constant-bias assumption in this setting is debatable. As noted above, the 2009 reform was progressive, so the two groups being compared may differ in unobserved characteristics. In such a setting, it is difficult to justify exact constancy on economic grounds; Appendix D of the OA provides an economic model illustrating this point, building on \cite{Fudenberg_Levine:2022AEJmicro}.

\begin{figure}[t]
    \centering
    \begin{subfigure}[b]{0.49\textwidth}
        \centering
        \includegraphics[width=\linewidth]{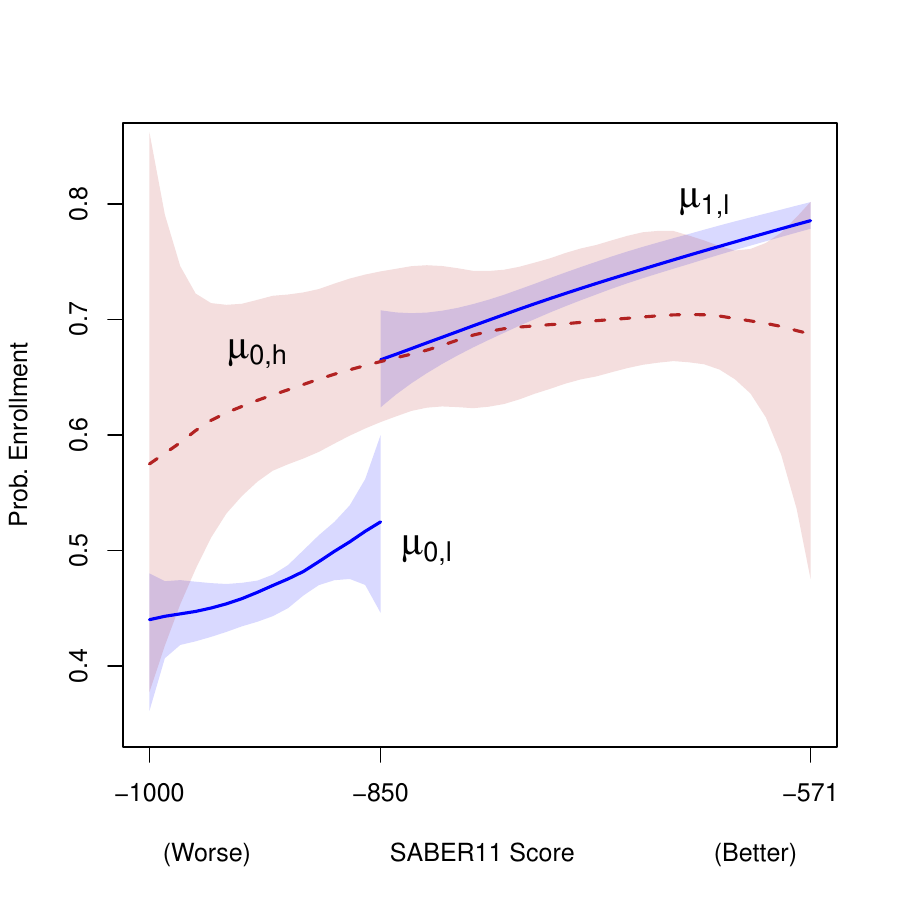}
        \caption{Observed Regression Functions}
        \label{fig:regs}
    \end{subfigure}
    \hfill
    \begin{subfigure}[b]{0.49\textwidth}
        \centering
        \includegraphics[width=\linewidth]{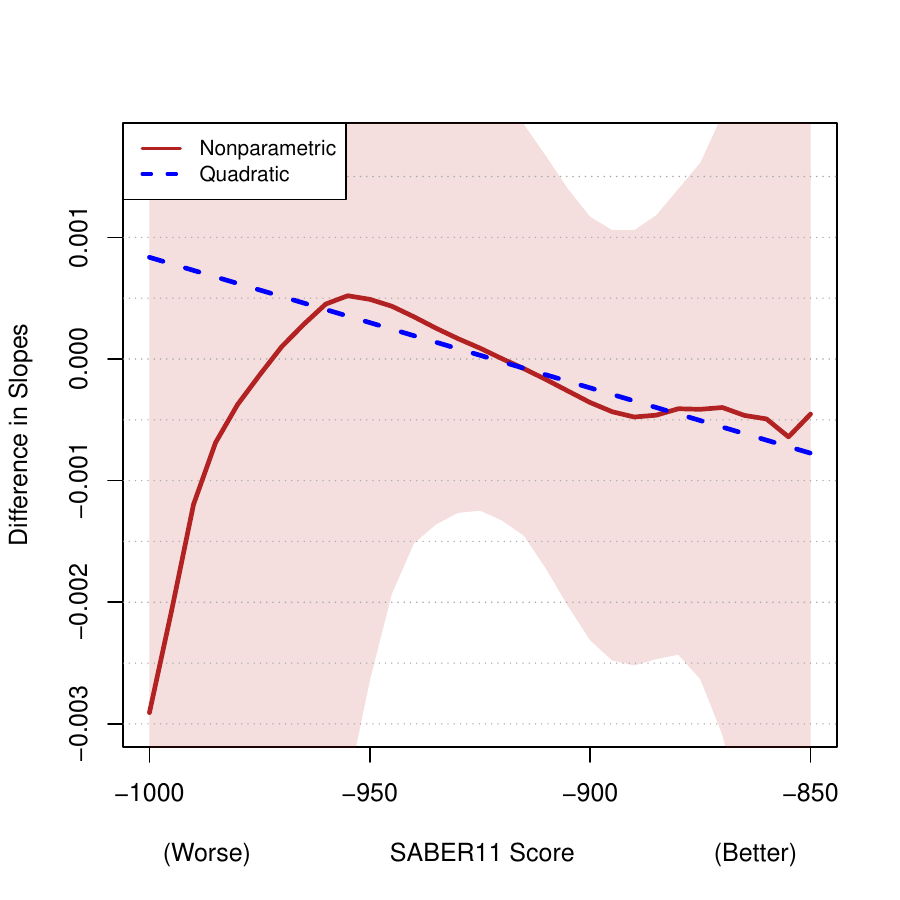}
        \caption{Derivative of the Bias Function}
        \label{fig:bias}
    \end{subfigure}
    \caption{Visual Assessment of Identifying Assumptions}
    \label{fig:assess}

    \begin{flushleft}
    \footnotesize
    \renewcommand{\baselineskip}{11pt}
    \textbf{Note:} The (total) sample size is $n=1{,}365$.
    Local polynomial estimates are reported.
    Shaded regions indicate the RBC confidence intervals.
    \end{flushleft}
\end{figure}

That said, abandoning constant-bias-based extrapolation altogether may be overly pessimistic. As Figure~\ref{fig:assess} suggests, although exact constancy is not obvious, the behavior of $\mu_\ell$ and $\mu_h$ over the region $x\leq \ell$ does not appear to differ dramatically. Moreover, \cite{Cattaneo_etal:2021JASA_extrapolating} reports that the constant-bias restriction cannot be statistically rejected over this region. These observations suggest that treating constant-bias as a benchmark and allowing for smooth deviations from it through our bounding approach provides a reasonable middle ground between imposing exact constant-bias and dispensing with it entirely.

Implementing this approach requires specifying the Lipschitz constant $B_h$. 
We set $B_h=0.0010$ as our baseline value and also examine sensitivity to $B_h \in [0.0001, 0.0030]$. The baseline choice of $B_h=0.0010$ allows the enrollment probability to deviate from the constant-bias prediction by up to $0.10$ (10 percentage points) for a 100-point change in the SABER 11 score, which corresponds to approximately 10\% of the score range. 
This appears to allow for a moderately large but still plausible violation of the constant-bias restriction. We assess the plausibility of this choice more directly in Section~\ref{subsubsec:assess} below.

\subsubsection{Testing Identifying Conditions}\label{subsubsec:assess}

A visual assessment of Figure~\ref{fig:regs} suggests RV- and group-monotonicity conditions are not refuted.
The plausibility of RV monotonicity can also be assessed indirectly by testing whether $\mu_{\ell}$ is monotonically increasing over $[-1000,-850)$ and $\mu_{h}$ is monotonically increasing over $[-1000,-571)$. We employ the monotonicity test of \cite{Hall_Heckman:2000}, which does not reject the null of monotonicity (with $p$-values of $0.240$ and $0.776$, respectively).

The group-monotonicity assumption has a directly testable implication at the lower cutoff. Under Assumption \ref{assumption:c-monotonicity}(i), $\lim_{x\uparrow\ell}\mu_{\ell}(x)-\mu_h(\ell)\leq 0$.
We assess this inequality using RBC local linear estimators. The resulting 95\% one-sided confidence interval for $\lim_{x\uparrow\ell}\mu_{\ell}(x)-\mu_h(\ell)$ is $(-\infty,-0.069]$, providing no evidence against the group monotonicity at the cutoff.

The smooth-bias condition is assessed in Figure~\ref{fig:bias}, where we estimate the slope of the bias function, $\delta_h^\prime(x)$, over $[-1000,-850)$ using an RBC local quadratic estimator and a parametric global quadratic specification. Under both specifications, the estimated derivative lies within $[-0.001,0.001]$ for most values of the running variable, which supports the baseline choice of $B_h = 0.001$.
The nonparametric fit indicates that the derivative is larger in absolute value near the lower boundary of the support, although this pattern may reflect sampling variability, as indicated by the wide confidence intervals in this region. We therefore report the sensitivity of our main results to alternative choices of $B_h$ up to $0.003$.

\subsection{Empirical Results}

\begin{figure}[t]
    \centering
    \includegraphics[width=0.75\linewidth]{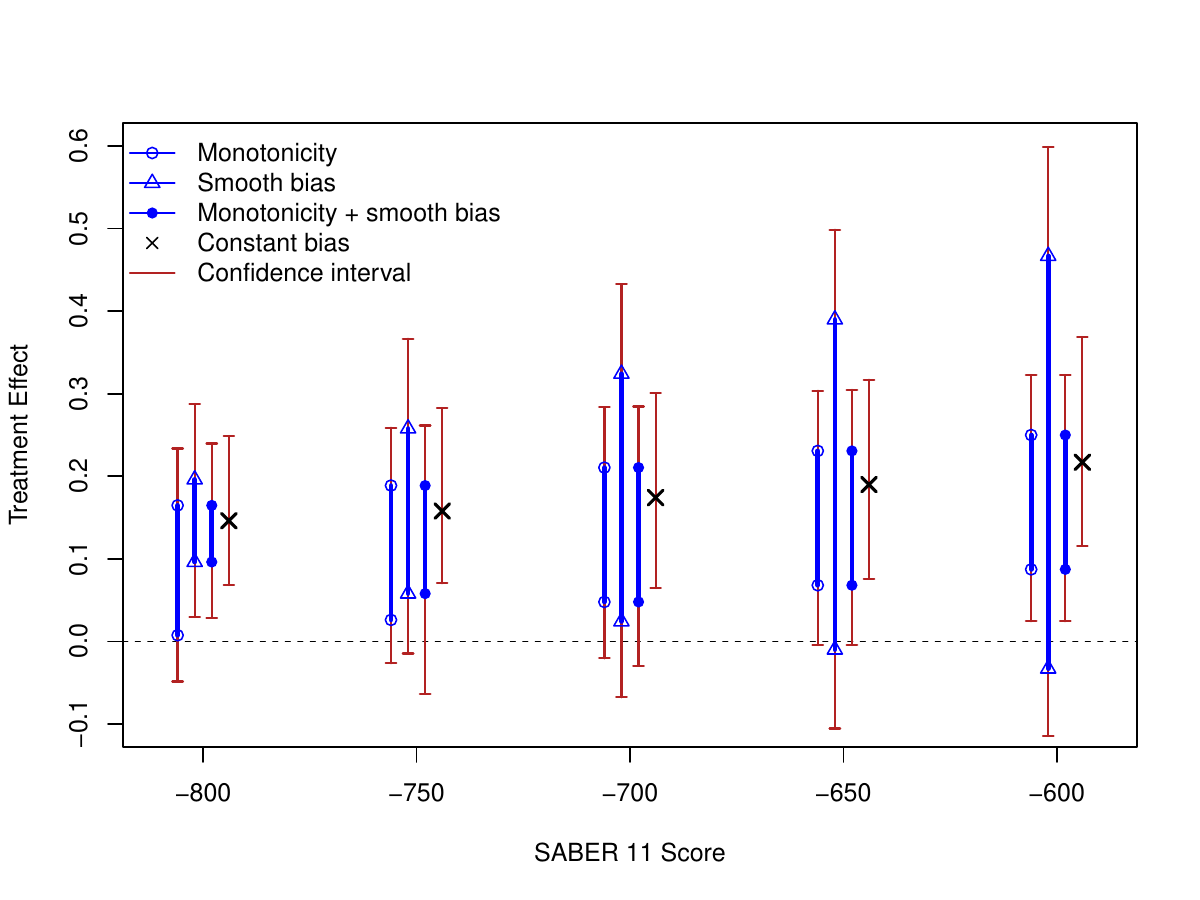}
    \caption{Treatment Effects of ACCES Away From the Cutoff}
    \label{fig:emp-bounds}

    \begin{flushleft}
    \footnotesize
    \renewcommand{\baselineskip}{11pt}
    \textbf{Note:} Estimated sharp bounds ($\mathcal{B}_{\mathtt{M1}}$, $\mathcal{B}_{\mathtt{S}}$, and ${\mathcal{B}}_{\mathtt{MS1}}$) are reported. 
    For the monotonicity-only and joint cases, the 95\% confidence intervals are based on the outer sets $\widetilde{\mathcal{B}}_{\mathtt{M1}}$ and $\widetilde{\mathcal{B}}_{\mathtt{M1}}\cap\mathcal{B}_{\mathtt{S}}$, respectively. 
    \end{flushleft}
\end{figure}

We are now in a position to present the main estimation results. Figure~\ref{fig:emp-bounds} reports the estimated bounds on the treatment effect $\tau_{\ell}(\bar x)$ for $\bar x \in \{-800,-750,\ldots,-600\}$. For point estimation, we report the sharp bounds, whereas the confidence intervals are constructed based on an outer set by inverting the hybrid test of \cite{Andrews_etal:2023}.

Several important observations emerge. First, the monotonicity restriction provides nontrivial identifying power. Across all five evaluation points, the bounds $\widehat{\mathcal{B}}_{\mathtt{M1}}(\bar x)$ consistently imply a positive treatment effect. 
Second, particularly near the lower cutoff, the smooth-bias assumption provides substantial additional identifying power. At $\bar x=-800$, the lower bound of $\widehat{\mathcal{B}}_{\mathtt{S}}(\bar x)$ is higher than that of $\widehat{\mathcal{B}}_{\mathtt{M1}}(\bar x)$. Consequently, the bounds under the joint restriction, $\widehat{\mathcal{B}}_{\mathtt{MS1}}(\bar x)$, are much tighter than those obtained under the monotonicity restriction alone.
Third, the identifying power of the smooth-bias restriction alone gradually diminishes as the evaluation point moves farther away from the lower cutoff. Nevertheless, the monotonicity restriction continues to provide identifying power away from the cutoff, so that the joint bounds $\widehat{\mathcal{B}}_{\mathtt{MS1}}(\bar x)$ remain informative even at more distant evaluation points.
Finally, the estimated sharp bounds coincide with the intersection bounds, $\widehat{\mathcal{B}}_{\mathtt{MS1}}(\bar x) = \widehat{\widetilde{\mathcal{B}}}_{\mathtt{M1}}\cap\widehat{\mathcal{B}}_{\mathtt{S}}$. This equality follows because, as discussed in Section~\ref{subsec:smoothbias}, $\widehat{\mu}_h^\prime(x)-B_h<0$ throughout $x\in[-850,-571)$ (not reported; see the replication package).

\begin{figure}[t]
    \centering
    \begin{subfigure}[b]{0.49\textwidth}
        \centering
        \includegraphics[width=\linewidth]{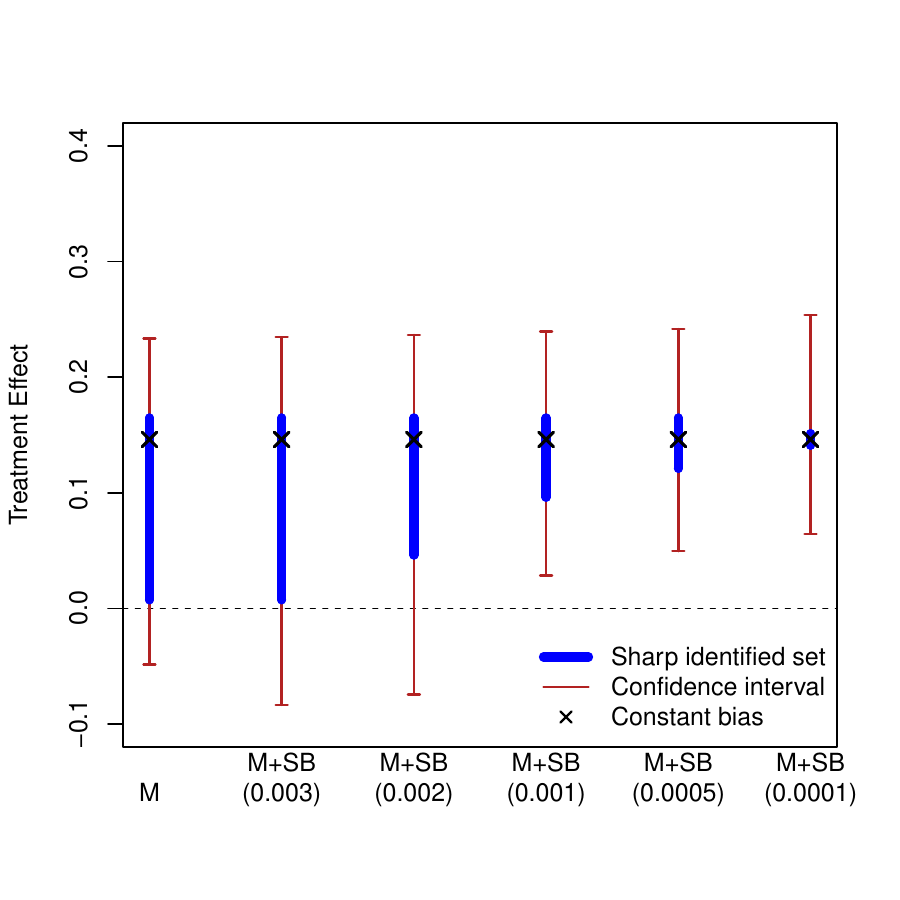}
        \caption{$\bar x = -800$}
        \label{fig:emp-800}
    \end{subfigure}
    \hfill
    \begin{subfigure}[b]{0.49\textwidth}
        \centering
        \includegraphics[width=\linewidth]{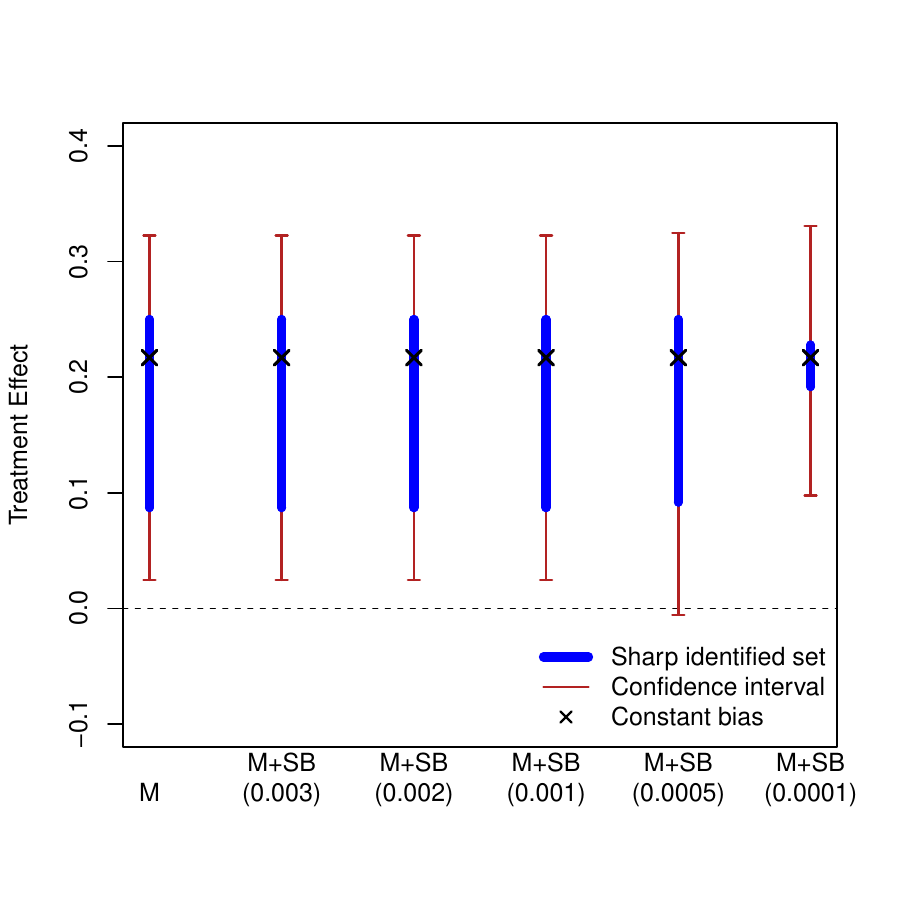}
        \caption{$\bar x = -600$}
        \label{fig:emp-600}
    \end{subfigure}
    \caption{Comparing Identifying Power}
    \label{fig:emp-power}

    \begin{flushleft}
    \footnotesize
    \renewcommand{\baselineskip}{11pt}
    \textbf{Note:} ``M'' denotes the monotonicity-only bounds, whereas ``M+SB'' denotes the bounds under the joint restriction. The value of the smoothness constant $B_h$ is reported in parentheses.
    \end{flushleft}
\end{figure}

Figure~\ref{fig:emp-power} examines the sensitivity of the identified sets to alternative choices of $B_h$, focusing on $\bar x=-800$, the evaluation point closest to the lower cutoff, and $\bar x=-600$, the one farthest from it.
Several findings emerge. First, at $\bar x=-800$, increasing $B_h$ from $0.001$ to $0.002$ and $0.003$ widens the joint bounds $\widehat{\mathcal{B}}_{\mathtt{MS1}}(\bar x)$. Further increases in $B_h$, however, do not lead to additional widening, because the monotonicity restrictions become binding and prevent the joint identified set from expanding beyond the monotonicity-based bounds. A similar pattern arises at $\bar x=-600$.
Second, at $\bar x=-800$, reducing $B_h$ from $0.001$ to $0.0005$ further tightens the joint bounds. At $\bar x=-600$, by contrast, the same reduction leaves the length of the identified set unchanged. Thus, smoothness restrictions play a relatively more important identifying role near the cutoff, whereas farther away their marginal identifying contribution diminishes.
That said, as $B_h$ is reduced further, the smoothness restriction recovers a strong identifying power, and the joint identified sets move toward the point-identified value implied by the constant-bias assumption of \cite{Cattaneo_etal:2021JASA_extrapolating}, corresponding to the limiting case $B_h=0$.

Taken together, Figures~\ref{fig:emp-bounds} and~\ref{fig:emp-power} illustrate the distinct and complementary identifying roles of the monotonicity and smoothness restrictions. The smoothness restriction is particularly informative near the cutoff, but its identifying power declines with the distance from the cutoff unless stronger smoothness is imposed. The monotonicity restriction does not exhibit the same systematic relationship with distance and may therefore retain meaningful identifying power farther from the cutoff. These findings highlight the practical value of combining the two restrictions when studying treatment-effect heterogeneity in multi-cutoff RD designs.

\section{Conclusion}\label{sec:concl}

This paper studied the identification of treatment effects beyond the cutoff in RD designs. We used monotonicity restrictions as a credible baseline and established sharp identification results for both canonical and multi-cutoff settings. We further investigated the additional identifying power provided by smoothness restrictions.
The resulting bounds offer an informative tool for assessing treatment-effect heterogeneity across values of the running variable and thereby help address the limited external validity of conventional RD designs. We illustrated the practical value of our approach through an empirical application, which highlighted the distinct and complementary identifying roles of monotonicity and smoothness restrictions.

\begin{singlespace}
\bibliographystyle{apalike}
\bibliography{refs}
\end{singlespace}

\newpage

\begin{center}
    \LARGE Online Appendices for \\
    ``Bounds on Extrapolated Treatment Effects in Regression Discontinuity Designs''
\end{center}
\begin{center}
    \large Yuta Okamoto \& Yuuki Ozaki
\end{center}

\appendix

\section{Supplement to the Main Results}\label{sec:proofs}

\subsection{Omitted Identification Results}

\begin{theorem}[Theorem \ref{thm:main1}, Continued]\label{thm:main1-OA}
    Let $\mathcal{C}=\{\ell,h\}$ and $\bar{x}\in(\ell,h)$.
    \begin{description}
        \item[(i)$^\prime$] Under Assumptions \ref{assumption:continuity},
        \ref{assumption:x-monotonicity}(ii), and
        \ref{assumption:c-monotonicity}(ii), the sharp identified set of
        $\tau_\ell(\bar{x})$ is given by
        \begin{align*}
            \bigg[
            \mu_\ell(\bar x)
            -
            \lim_{x\uparrow\ell}\mu_\ell(x),
            \quad
            \mu_\ell(\bar x)
            -
            \sup_{x\in[\bar x,h)}\mu_h(x)
            \bigg].
        \end{align*}
        
        \item[(ii)$^\prime$] Under Assumptions \ref{assumption:continuity},
        \ref{assumption:x-monotonicity}(ii), and
        \ref{assumption:c-monotonicity}(i), the sharp identified set of
        $\tau_\ell(\bar{x})$ is given by
        \begin{align*}
            \bigg[
            \left(
            \mu_\ell(\bar x)
            -
            \inf_{x\in[\ell,\bar x]}\mu_h(x)
            \right)
            \vee
            \left(
            \mu_\ell(\bar x)
            -
            \lim_{x\uparrow\ell}\mu_\ell(x)
            \right),
            \quad
            \mu_\ell(\bar x)-M_1
            \bigg],
        \end{align*}
        provided that $Y_i(0)\in[M_1,M_2]$ almost surely.
    \end{description}
\end{theorem}

\begin{theorem}[Theorem \ref{thm:main2}, Continued]\label{thm:main2-OA}
    Let $\mathcal{C}=\{\ell,h\}$ and $\bar{x}\in(\ell,h)$.
    \begin{description}
        \item[(ii)$^\prime$] Under the assumptions corresponding to case (i)$^\prime$ of Theorem~\ref{thm:main1-OA}, together with Assumption~\ref{assumption:lipchitz-bias}, the sharp identified set of $\tau_\ell(\bar x)$ is given by
        \begin{align*}
            \bigg[
            &
            \mu_\ell(\bar x) - \lim_{x\uparrow\ell}\mu_\ell(x) + V_{B_h}^{-}(\ell,\bar x),\\
            &\quad
            \left\{\tau_\ell^{\mathtt{CB}}(\bar x) + B_h(\bar x-\ell)\right\}
            \wedge
            \inf_{x\in[\bar x,h)} \left\{\mu_\ell(\bar x) - \mu_h(x) - V_{B_h}^{-}(\bar x,x)\right\}
            \bigg],
        \end{align*}
        where
        \begin{align*}
            V_{B_h}^{-}(s,t)
            :=
            \sup_{\substack{K\geq1,\\ s=x_0 \le x_1 \le \cdots \le x_K=t}}
            \sum_{k=1}^K
            \left[\mu_h(x_{k-1}) - \mu_h(x_k) - B_h(x_k-x_{k-1})
            \right]_+.
        \end{align*}
        
        \item[(iii)$^\prime$] Under the assumptions corresponding to case (ii)$^\prime$ of Theorem~\ref{thm:main1-OA}, together with Assumption~\ref{assumption:lipchitz-bias}, the sharp identified set of $\tau_\ell(\bar x)$ is given by
        \begin{align*}
            \bigg[
            &
            \left\{\mu_\ell(\bar x) - \lim_{x\uparrow\ell}\mu_\ell(x) + V_{B_h}^{-}(\ell,\bar x)\right\}
            \vee
            \sup_{x\in[\ell,\bar x]}\left\{\mu_\ell(\bar x) - \mu_h(x) + V_{B_h}^{-}(x,\bar x)\right\},\\
            &\quad
            \left\{\tau_\ell^{\mathtt{CB}}(\bar x) + B_h(\bar x-\ell)\right\}
            \wedge
            \inf_{x\in[\bar x,h)} \left\{\mu_\ell(\bar x) - M_1 - V_{B_h}^{-}(\bar x,x)\right\}
            \bigg],
        \end{align*}
        provided that $Y_i(0)\in[M_1,M_2]$ almost surely.
    \end{description}
\end{theorem}

\subsection{Proofs of Theorems}

\begin{proof}[Proof of Theorem \ref{thm:main1}]
    We begin by proving parts (i) and (ii).
    We first establish validity of the stated bounds.
    In case (i), RV-monotonicity implies $\mu_{0,\ell}(\bar x) \ge \mu_{0,\ell}(\ell) = \lim_{x\uparrow\ell}\mu_\ell(x)$.
    Moreover, for every $x\in[\bar x,h)$, RV-monotonicity and group-monotonicity imply $\mu_{0,\ell}(\bar x) \le \mu_{0,\ell}(x) \le \mu_{0,h}(x) = \mu_h(x)$.
    It then follows that $\mu_{0,\ell}(\bar x) \le \inf_{x\in[\bar x,h)}\mu_h(x)$.
    Hence,
    \begin{align*}
        \lim_{x\uparrow\ell}\mu_\ell(x)
        \leq
        \mu_{0,\ell}(\bar x)
        \leq
        \inf_{x\in[\bar x,h)}\mu_h(x).
    \end{align*}
    Since $\tau_\ell(\bar x)=\mu_\ell(\bar x)-\mu_{0,\ell}(\bar x)$, this yields $\mathcal B_{\mathtt{M1}}(\bar x)$.
    In case (ii), RV-monotonicity implies $\mu_{0,\ell}(\bar x) \ge \lim_{x\uparrow\ell}\mu_\ell(x)$.
    Moreover, for every $x\in[\ell,\bar x]$, RV-monotonicity and group-monotonicity imply $\mu_{0,\ell}(\bar x)\geq\mu_{0,\ell}(x)\geq\mu_{0,h}(x)=\mu_h(x)$.
    We therefore have that $\mu_{0,\ell}(\bar x) \ge \sup_{x\in[\ell,\bar x]}\mu_h(x)$.
    Together with the support restriction,
    \begin{align*}
        \left(
        \lim_{x\uparrow\ell}\mu_\ell(x)
        \right)
        \vee
        \left(
        \sup_{x\in[\ell,\bar x]}\mu_h(x)
        \right)
        \leq
        \mu_{0,\ell}(\bar x)
        \leq
        M_2.
    \end{align*}
    This is equivalent to the bounds in $\mathcal B_{\mathtt{M2}}(\bar x)$.

    It remains only to establish sharpness.
    Fix any $b$ in the identified set stated in the corresponding part of the theorem, and let $a:= \mu_\ell(\bar x)-b$.
    In case (i), define, for $x\in[\ell,\bar x]$,
    \begin{align*}
        f_a(x)
        :=
        \min\left\{
        \frac{
            a-\lim_{z\uparrow\ell}\mu_\ell(z)
        }{
            \bar x-\ell
        }(x-\ell)
        +
        \lim_{z\uparrow\ell}\mu_\ell(z),
        \quad
        \inf_{z\in[x,\bar x]}\mu_h(z)
        \right\}.
    \end{align*}
    Both functions inside the minimum are weakly increasing in $x$, and hence so is $f_a$.
    Compatibility of the maintained restrictions implies $\lim_{z\uparrow\ell}\mu_\ell(z) \le \inf_{z\in[\ell,\bar x]}\mu_h(z)$, while admissibility (attainability) of $a$ implies $a \le \inf_{z\in[\bar x,h)}\mu_h(z) \le \mu_h(\bar x)$.
    It then follows that $f_a(\ell) = \lim_{z\uparrow\ell}\mu_\ell(z)$ and $f_a(\bar x)=a$.
    Moreover, $f_a(x) \le \inf_{z\in[x,\bar x]}\mu_h(z) \le \mu_h(x)$ for every $x\in[\ell,\bar x]$.
    Here, define a completion of the unobserved conditional mean function by
    \begin{align*}
        \mu_{0,\ell}(x)
        =
        \begin{cases}
            \mu_\ell(x),
            & \text{if }x\in(-\infty,\ell)\cap\mathcal X,\\
            f_a(x),
            & \text{if }x\in[\ell,\bar x],\\
            a,
            & \text{if }x\in(\bar x,\infty)\cap\mathcal X.
        \end{cases}
    \end{align*}
    The resulting function is continuous and weakly increasing for $x\ge \ell$.
    For $x\in[\ell,\bar x]$, the preceding construction gives $\mu_{0,\ell}(x)\leq\mu_h(x)$.
    For $x\in(\bar x,h)$, admissibility of $a$ gives $\mu_{0,\ell}(x) = a \le \inf_{z\in[\bar x,h)}\mu_h(z) \le \mu_h(x)$. By continuity, the same ordering holds at $h$. Hence, both monotonicity restrictions are satisfied.

    In case (ii), define, for $x\in[\ell,\bar x]$,
    \begin{align*}
        f_a(x)
        :=
        \max\left\{
        \frac{a-\lim_{z\uparrow\ell}\mu_\ell(z)}{\bar x-\ell}(x-\ell)
        +
        \lim_{z\uparrow\ell}\mu_\ell(z),
        \quad
        \sup_{z\in[\ell,x]}\mu_h(z)
        \right\}.
    \end{align*}
    Both functions inside the maximum are weakly increasing in $x$,
    and hence so is $f_a$.
    Compatibility of the maintained restrictions implies $\mu_h(\ell)\leq\lim_{z\uparrow\ell}\mu_\ell(z)$, while admissibility of $a$ implies $a \ge \sup_{z\in[\ell,\bar x]}\mu_h(z)$.
    Hence, $f_a(\ell) = \lim_{z\uparrow\ell}\mu_\ell(z)$, $f_a(\bar x)=a$, and $f_a(x) \ge \sup_{z\in[\ell,x]}\mu_h(z) \ge \mu_h(x)$ for every $x\in[\ell,\bar x]$.
    Define
    \begin{align*}
        \mu_{0,\ell}(x)
        :=
        \begin{cases}
            \mu_\ell(x),
            & \text{if }x\in(-\infty,\ell)\cap\mathcal X,\\
            f_a(x),
            & \text{if }x\in[\ell,\bar x].
        \end{cases}
    \end{align*}
    and for $x\in(\bar x, h)$,
    \begin{align*}
        \mu_{0,\ell}(x)
        :=
        \max\left\{
            a,\,
            \sup_{z\in[\bar x,x]}\mu_h(z)
        \right\}.
    \end{align*}
    At $h$, set
    \begin{align*}
        \mu_{0,\ell}(h)
        :=
        \max\left\{
            a,\,
            \sup_{z\in[\bar x,h)}\mu_h(z)
        \right\},
    \end{align*}
    and keep $\mu_{0,\ell}$ constant at this value for $x>h$.
    This completion is continuous and weakly increasing.
    It also satisfies $\mu_{0,\ell}(x)\geq\mu_h(x)$ for every $x\in[\ell,h)$, and by continuity the group ordering also holds at $h$.
    Moreover, since $Y_i(0)\in[M_1,M_2]$ almost surely, $\mu_h(x)\leq M_2$ for $x<h$, while admissibility implies $a\leq M_2$.
    Hence the constructed mean remains in $[M_1,M_2]$.

    In both cases, the construction changes $\mu_{0,\ell}(x)$ only for $x\geq\ell$ and therefore leaves the distribution of observed outcomes unchanged.
    On this unobserved region, one can choose a conditional distribution of $Y_i(0)$ having the constructed mean and, whenever the support restriction is imposed, support contained in $[M_1,M_2]$, while leaving all observed conditional distributions unchanged. Therefore every admissible value $a$ is attainable, and the identified sets in parts (i) and (ii) are sharp. Parts (i)$^\prime$ and (ii)$^\prime$ of Theorem~\ref{thm:main1-OA} are analogous.

    Finally, we show part (iii).
    Since $\mu_{0,h}(x)=\mu_h(x)$ for $x<h$, in cases (i)--(ii) the additional restriction implies $\inf_{x\in[\bar x,h)}\mu_h(x) = \sup_{x\in[\ell,\bar x]}\mu_h(x) = \mu_h(\bar x)$, whereas in cases (i)$^\prime$--(ii)$^\prime$ it implies $\inf_{x\in[\ell,\bar x]}\mu_h(x) = \sup_{x\in[\bar x,h)}\mu_h(x) = \mu_h(\bar x)$.
    Substituting these equalities into parts (i)--(ii)$^\prime$ yields exactly the intersections of the corresponding bounds in Propositions~\ref{prop:rv} and~\ref{prop:group}. Sharpness follows from the constructions in parts (i)--(ii)$^\prime$, which leave $\mu_{0,h}$ unchanged.
\end{proof}

\begin{proof}[Proof of Theorem \ref{thm:main2}]
\textit{Proof of part (i).} 
    Since $\mu_{0,h}-\mu_{0,\ell}\in \mathcal L(B_h)$, the Lipschitz restriction together with the continuity assumption implies, for $x\in[\ell,\bar x]$,
    \begin{align*}
        \mu_h(\ell) - \lim_{x\uparrow \ell}\mu_{\ell}(x) - B_h ({x}-\ell) \le 
        \mu_h(x) - \mu_{0,\ell}(x) \le
        \mu_h(\ell) - \lim_{x\uparrow \ell}\mu_{\ell}(x) + B_h ({x}-\ell),
    \end{align*}
    which is equivalent to $\tau_\ell(x) \in [\tau_\ell^{\mathtt{CB}}(x)-B_h(x-\ell),\, \tau_\ell^{\mathtt{CB}}(x)+B_h(x-\ell)]$ for $x\in[\ell,\bar x]$, where $\tau_\ell^{\mathtt{CB}}(x) = \mu_\ell(x)-\mu_h(x)+\mu_h(\ell)-\lim_{z\uparrow\ell}\mu_\ell(z)$.
    Then, $\mathcal B_{\mathtt S}(\bar x) := [\tau_\ell^{\mathtt{CB}}(\bar x)-B_h(\bar x-\ell), \tau_\ell^{\mathtt{CB}}(\bar x)+B_h(\bar x-\ell)]$ is a valid identified set.
    We next show that $\mathcal B_{\mathtt S}(\bar x)$ is sharp.
    Take any $b\in\mathcal B_{\mathtt S}(\bar x)$ and let $a:=b-\tau_\ell^{\mathtt{CB}}(\bar x)$. By construction, $|a|\leq B_h(\bar x-\ell)$.
    Consider the following completion of the bias function:
    \begin{align*}
        \widetilde\delta_h(x)
        =
        \begin{cases}
            \delta_h(x)
            & \text{if } x\leq\ell,\\
            \delta_h(\ell)
            +\dfrac{a}{\bar x-\ell}(x-\ell)
            & \text{if } x\in[\ell,\bar x],\\
            \delta_h(\ell)+a
            & \text{if } x\geq\bar x.
        \end{cases}
    \end{align*}
    Since $|a|/(\bar x-\ell)\leq B_h$, $\widetilde\delta_h \in \mathcal L(B_h)$.
    Define the corresponding counterfactual mean $\widetilde\mu_{0,\ell}(x) = \mu_{0,h}(x)-\widetilde\delta_h(x)$. This $\widetilde\mu_{0,\ell}$ agrees with the observed lower-cutoff regression function for $x<\ell$ and modifies only the counterfactual untreated mean for the lower-cutoff group when $x\geq\ell$. It therefore leaves the distribution of observables unchanged.
    Furthermore, at $x = \bar x$, the treatment effect is $\mu_\ell(\bar x)-\mu_h(\bar x)+\widetilde\delta_h(\bar x) = \tau_\ell^{\mathtt{CB}}(\bar x)+a = b$. Hence, every $b\in\mathcal B_{\mathtt S}(\bar x)$ is attainable, and $\mathcal B_{\mathtt S}(\bar x)$ is the sharp identified set under Assumptions \ref{assumption:continuity} and \ref{assumption:lipchitz-bias}.

\noindent\textit{Proof of part (ii).}
    We first establish validity of the stated bounds.
    For any $\ell\leq s \leq t<h$, the smooth-bias restriction imply
    \begin{align*}
        \mu_{0,\ell}(t)-\mu_{0,\ell}(s)
        &=
        \mu_h(t)-\mu_h(s) - \{\delta_h(t)-\delta_h(s)\}\\
        &\geq
        \mu_h(t) - \mu_h(s) - B_h(t-s),
    \end{align*}
    while RV-monotonicity also implies
    $\mu_{0,\ell}(t)-\mu_{0,\ell}(s)\geq0$.
    Hence,
    \begin{align*}
        \mu_{0,\ell}(t)-\mu_{0,\ell}(s)
        \geq
        \left[\mu_h(t) - \mu_h(s) - B_h(t-s)\right]_+.
    \end{align*}
    Applying this inequality along any partition $s=x_0 \le x_1 \le \cdots \le x_K=t$ and summing gives $\mu_{0,\ell}(t)-\mu_{0,\ell}(s) \ge \sum_{k=1}^K \left[\mu_h(x_k) - \mu_h(x_{k-1}) - B_h(x_k-x_{k-1})\right]_+$.
    Taking the supremum over all such partitions yields
    \begin{align}
        \mu_{0,\ell}(t)-\mu_{0,\ell}(s)
        \geq
        V_{B_h}^{+}(s,t).\label{eq:V}
    \end{align}
    Setting $s=\ell$ and $t=\bar x$ in \eqref{eq:V}, continuity implies $\mu_{0,\ell}(\bar x) \geq \lim_{x\uparrow\ell}\mu_\ell(x) + V_{B_h}^{+}(\ell,\bar x)$.
    Therefore,
    \begin{align}
        \tau_\ell(\bar x)
        \leq
        \mu_\ell(\bar x) - \lim_{x\uparrow\ell}\mu_\ell(x) - V_{B_h}^{+}(\ell,\bar x).\label{eq:MS1-upper}
    \end{align}
    Next, for any $x\in[\bar x,h)$, \eqref{eq:V} and group-monotonicity imply $\mu_h(x) \geq \mu_{0,\ell}(x) \ge \mu_{0,\ell}(\bar x)+V_{B_h}^{+}(\bar x,x)$.
    Then we have $\mu_{0,\ell}(\bar x) \leq \mu_h(x)-V_{B_h}^{+}(\bar x,x)$ for every $x\in[\bar x,h)$, which implies
    \begin{align}
        \tau_\ell(\bar x)
        \geq
        \sup_{x\in[\bar x,h)} \left\{\mu_\ell(\bar x) - \mu_h(x) + V_{B_h}^{+}(\bar x,x)\right\}.\label{eq:MS1-lower-mon}
    \end{align}
    Finally, the smooth-bias restriction between $\ell$ and $\bar x$ implies $\delta_h(\bar x) \geq \delta_h(\ell)-B_h(\bar x-\ell)$.
    Since $\delta_h(\ell) = \mu_h(\ell)-\lim_{x\uparrow\ell}\mu_\ell(x)$ and $\tau_\ell(\bar x) = \mu_\ell(\bar x)-\mu_h(\bar x)+\delta_h(\bar x)$, it follows that
    \begin{align}
        \tau_\ell(\bar x)
        \geq
        \tau_\ell^{\mathtt{CB}}(\bar x)
        -
        B_h(\bar x-\ell).
        \label{eq:MS1-lower-smooth}
    \end{align}
    Combining \eqref{eq:MS1-upper}--\eqref{eq:MS1-lower-smooth} establishes validity of $\mathcal B_{\mathtt{MS1}}(\bar x)$.

    It remains to establish sharpness.
    Fix any $b\in\mathcal B_{\mathtt{MS1}}(\bar x)$ and let $a:=\mu_\ell(\bar x)-b$.
    For $x\in[\ell,\bar x]$, define
    \begin{align}
        f_a(x)
        :=
        \max\Bigg\{
        \lim_{z\uparrow\ell}\mu_\ell(z) + V_{B_h}^{+}(\ell,x),\quad
        a - \mu_h(\bar x) + \mu_h(x) - B_h(\bar x-x)
        \Bigg\},\label{eq:fa-left}
    \end{align}
    and, for $x\in[\bar x,h)$, define
    \begin{align}
        f_a(x) := a + V_{B_h}^{+}(\bar x,x).\label{eq:fa-right}
    \end{align}
    We verify that this construction satisfies the maintained restrictions.
    First, admissibility of $b$ and the upper endpoint of $\mathcal B_{\mathtt{MS1}}(\bar x)$ imply $a \ge \lim_{z\uparrow\ell}\mu_\ell(z) + V_{B_h}^{+}(\ell,\bar x)$. Moreover, the lower smooth-bias bound implies $a \le \lim_{z\uparrow\ell}\mu_\ell(z) + \mu_h(\bar x)-\mu_h(\ell) + B_h(\bar x-\ell)$.
    It then follows from \eqref{eq:fa-left} that $f_a(\ell) = \lim_{z\uparrow\ell}\mu_\ell(z)$ and $f_a(\bar x)=a$.
    Likewise, the other component of the lower endpoint of
    $\mathcal B_{\mathtt{MS1}}(\bar x)$ implies $a \le \mu_h(x)-V_{B_h}^{+}(\bar x,x)$ for every $x\in[\bar x,h)$.
    Hence \eqref{eq:fa-right} satisfies $f_a(x)\leq\mu_h(x)$ for $x\in[\bar x,h)$.
    
    To verify the remaining shape restrictions, note first that $x\mapsto V_{B_h}^{+}(s,x)$ is weakly increasing, and hence $x\mapsto \lim_{z\uparrow\ell}\mu_\ell(z) + V_{B_h}^{+}(\ell,x)$ is weakly increasing.
    Next, take arbitrary feasible counterfactual mean function $\mu_{0,\ell}^*$, which exists provided that the maintained assumptions are true. Then, we have $\mu_h(x) - \lim_{z\uparrow\ell}\mu_{\ell}(z) \ge \mu_{0,\ell}^*(x) - \lim_{z\uparrow\ell}\mu_{\ell}(z) = \mu_{0,\ell}^*(x) - \mu_{0,\ell}^*(\ell) \ge V_{B_h}^+(\ell, x)$, where the last inequality uses \eqref{eq:V}. Thus $\lim_{z\uparrow\ell}\mu_\ell(z) + V_{B_h}^{+}(\ell,x)$ lies weakly below $\mu_h(x)$. Finally, for $\ell\le x_1 \le x_2 < h$,
    \begin{align*}
        &\left\{\mu_h(x_2) - \lim_{z\uparrow\ell}\mu_\ell(z) - V_{B_h}^{+}(\ell,x_2)\right\} - \left\{\mu_h(x_1) - \lim_{z\uparrow\ell}\mu_\ell(z) - V_{B_h}^{+}(\ell,x_1)\right\}\\
        &=
        \mu_h(x_2) - \mu_h(x_1) + V_{B_h}^{+}(\ell,x_1) - V_{B_h}^{+}(\ell,x_2)
        =
        \mu_h(x_2) - \mu_h(x_1) - V_{B_h}^{+}(x_1,x_2),
    \end{align*}
    where we used the additivity of the positive variation: $V_{B_h}^{+}(s,t) = V_{B_h}^{+}(s,x) + V_{B_h}^{+}(x,t)$ for $s\leq x\leq t$.
    By definition of $V_{B_h}^{+}$, $V_{B_h}^{+}(x_1,x_2) \ge \mu_h(x_2) - \mu_h(x_1) - B_h(x_2 - x_1)$, so that $\mu_h(x_2) - \mu_h(x_1) - V_{B_h}^{+}(x_1,x_2) \le B_h(x_2 - x_1)$.
    Moreover, \eqref{eq:V} and smooth-bias imply that $\mu_h(x_2) - \mu_h(x_1) - V_{B_h}^{+}(x_1,x_2) \ge -B_h(x_2 - x_1)$. Therefore, $\mu_h(x) - \{\lim_{z\uparrow\ell}\mu_\ell(z) + V_{B_h}^{+}(\ell,x)\}$ is $B_h$-Lipschitz continuous on $[\ell, \bar x]$.
    
    The second function inside the maximum in \eqref{eq:fa-left} can similarly be shown to be weakly increasing, lies weakly below $\mu_h(x)$, and its difference from $\mu_h(x)$ has slope $-B_h$.
    Therefore $f_a$ in \eqref{eq:fa-left} is weakly increasing, satisfies $f_a(x)\leq\mu_h(x)$, and $x\mapsto\mu_h(x)-f_a(x)$ is $B_h$-Lipschitz continuous on $[\ell,\bar x]$.
    Similar manipulation shows the function in \eqref{eq:fa-right} is weakly increasing and $x\mapsto\mu_h(x)-f_a(x)$ is $B_h$-Lipschitz on $[\bar x,h)$.
    Since the two pieces agree at $\bar x$, the same properties hold over $[\ell,h)$.

    We may therefore define
    \begin{align*}
        \mu_{0,\ell}(x)
        =
        \begin{cases}
            \mu_\ell(x),
            & \text{if }x\in(-\infty,\ell)\cap\mathcal X,\\
            f_a(x),
            & \text{if }x\in[\ell,h)\cap\mathcal X,
        \end{cases}
    \end{align*}
    and extend $\mu_{0,\ell}$ and $\mu_{0,h}$ beyond $h$ continuously so that both are weakly increasing and their difference remains $B_h$-Lipschitz; for example, they may both be held constant from their respective limiting values at $h$.
    The resulting completion satisfies continuity, RV-monotonicity, group-monotonicity, and Assumption \ref{assumption:lipchitz-bias}, while attaining $\mu_{0,\ell}(\bar x)=a$.
    Finally, the construction modifies only unobserved potential-outcome regressions and therefore leaves the distribution of observed outcomes unchanged.
    Hence every $b\in\mathcal B_{\mathtt{MS1}}(\bar x)$ is attainable, proving sharpness.

\noindent\textit{Proof of part (iii).}
    We first establish validity.
    As shown in part (ii), RV-monotonicity and the smooth-bias restriction imply \eqref{eq:V}. Hence, $\mu_{0,\ell}(\bar x) \ge \lim_{x\uparrow\ell}\mu_\ell(x) + V_{B_h}^{+}(\ell,\bar x)$.
    Moreover, for every $x\in[\ell,\bar x]$, \eqref{eq:V} and group-monotonicity imply $\mu_{0,\ell}(\bar x) \ge \mu_{0,\ell}(x) + V_{B_h}^{+}(x,\bar x) \ge \mu_h(x) + V_{B_h}^{+}(x,\bar x)$.
    It then follows that
    \begin{align*}
        \mu_{0,\ell}(\bar x)
        \geq
        \left\{
        \lim_{x\uparrow\ell}\mu_\ell(x)
        +
        V_{B_h}^{+}(\ell,\bar x)
        \right\}
        \vee
        \sup_{x\in[\ell,\bar x]}
        \left\{
        \mu_h(x)
        +
        V_{B_h}^{+}(x,\bar x)
        \right\}.
    \end{align*}
    On the other hand, we have seen that the smooth-bias restriction between $\ell$ and $\bar x$ gives $\mu_{0,\ell}(\bar x) \le \lim_{x\uparrow\ell}\mu_\ell(x) + \mu_h(\bar x)-\mu_h(\ell) + B_h(\bar x-\ell)$.
    Also, for every $x\in[\bar x,h)$, \eqref{eq:V} and the support restriction imply $M_2 \ge \mu_{0,\ell}(x) \ge \mu_{0,\ell}(\bar x)+V_{B_h}^{+}(\bar x,x)$.
    Hence
    \begin{align*}
        \mu_{0,\ell}(\bar x)
        \leq
        \left\{
        \lim_{x\uparrow\ell}\mu_\ell(x)
        +
        \mu_h(\bar x)-\mu_h(\ell)
        +
        B_h(\bar x-\ell)
        \right\}
        \wedge
        \inf_{x\in[\bar x,h)}
        \left\{
        M_2-V_{B_h}^{+}(\bar x,x)
        \right\}.
    \end{align*}
    These inequalities establish validity of $\mathcal B_{\mathtt{MS2}}(\bar x)$.

    It remains to establish sharpness.
    Fix any $b\in\mathcal B_{\mathtt{MS2}}(\bar x)$ and let $a:=\mu_\ell(\bar x)-b$.
    For $c\le x<h$, define
    \begin{align*}
        Q_c(x)
        :=
        \sup_{z\in[c,x]} \left\{ \mu_h(z)+V_{B_h}^{+}(z,x) \right\}.
    \end{align*}
    By additivity of $V_{B_h}^{+}$, for $c\leq s<t<h$,
    \begin{align*}
        Q_c(t)&= 
        \sup_{z\in[c,t]} \left\{ \mu_h(z)+V_{B_h}^{+}(z,t) \right\}
        \geq 
        \sup_{z\in[c,s]} \left\{ \mu_h(z)+V_{B_h}^{+}(z,t) \right\}\\
        &=
        \sup_{z\in[c,s]} \left\{ \mu_h(z)+V_{B_h}^{+}(z,s)+V_{B_h}^{+}(s,t) \right\}
        \geq Q_c(s)+V_{B_h}^{+}(s,t),
    \end{align*}
    and hence $Q_c$ is weakly increasing.
    We then establish an upper bound on the increment of $Q_c$.
    As shown in part (ii), feasibility of the maintained restrictions implies
    \begin{align*}
        V_{B_h}^{+}(x_1,x_2)
        \leq
        \mu_h(x_2)-\mu_h(x_1)+B_h(x_2-x_1)
    \end{align*}
    for any $x_1\leq x_2<h$.
    For any $z\in[c,s]$, additivity therefore gives
    \begin{align*}
        \mu_h(z)+V_{B_h}^{+}(z,t)
        =
        \mu_h(z)+V_{B_h}^{+}(z,s)+V_{B_h}^{+}(s,t)
        \leq
        Q_c(s)+\mu_h(t)-\mu_h(s)+B_h(t-s).
    \end{align*}
    For any $z\in[s,t]$, we similarly have
    \begin{align*}
        \mu_h(z)+V_{B_h}^{+}(z,t)
        &=
        \mu_h(t) + \mu_h(z) - \mu_h(t) +V_{B_h}^{+}(z,t) 
        \leq
        \mu_h(t)+B_h(t-z)\\
        &\leq
        \mu_h(t)+B_h(t-s)
        \leq
        Q_c(s)+\mu_h(t)-\mu_h(s)+B_h(t-s),
    \end{align*}
    where the last inequality follows from $Q_c(s)\geq\mu_h(s)$.
    Taking the supremum over $z\in[c,t]$ yields
    \begin{align*}
        Q_c(t)-Q_c(s) \leq \mu_h(t)-\mu_h(s)+B_h(t-s).
    \end{align*}
    On the other hand, $Q_c(t)-Q_c(s) \ge V_{B_h}^{+}(s,t) \ge \mu_h(t)-\mu_h(s)-B_h(t-s)$.
    Thus,
    \begin{align}
        \mu_h(t)-\mu_h(s)-B_h(t-s)
        \leq
        Q_c(t)-Q_c(s)
        \leq
        \mu_h(t)-\mu_h(s)+B_h(t-s).\label{eq:Q}
    \end{align}
    In particular, $Q_c-\mu_h$ is $B_h$-Lipschitz continuous, and hence $Q_c$ is continuous.
    
    Now, for $x\in[\ell,\bar x]$, define
    \begin{align*}
        f_a(x) :=
        \max\Bigg\{
        \lim_{z\uparrow\ell}\mu_\ell(z)
        +V_{B_h}^{+}(\ell,x),\,\,
        Q_\ell(x),\,\,
        a-\mu_h(\bar x)
        +\mu_h(x)
        -B_h(\bar x-x)
        \Bigg\},
    \end{align*}
    and, for $x\in[\bar x,h)$, define
    \begin{align*}
        f_a(x) :=
        \max\left\{
        a+V_{B_h}^{+}(\bar x,x),\,\,
        Q_{\bar x}(x)
        \right\}.
    \end{align*}
    We verify that this construction satisfies the maintained assumptions. 
    First, admissibility of $b$ implies $a \ge \lim_{z\uparrow\ell}\mu_\ell(z) + V_{B_h}^{+}(\ell,\bar x)$, $a \ge Q_\ell(\bar x)$, and $a \le \lim_{z\uparrow\ell}\mu_\ell(z) + \mu_h(\bar x)-\mu_h(\ell) + B_h(\bar x-\ell)$.
    Together with group-monotonicity at $\ell$, these inequalities imply that $f_a(\ell) = \lim_{z\uparrow\ell}\mu_\ell(z)$ and $f_a(\bar x)=a$.
    Moreover, $Q_{\bar x}(\bar x)=\mu_h(\bar x)\leq a$, so the two definitions agree at $\bar x$.

    As shown in part (ii), $x\mapsto \lim_{z\uparrow\ell}\mu_\ell(z) + V_{B_h}^{+}(\ell,x)$ and $x\mapsto a+V_{B_h}^{+}(\bar x,x)$ are weakly increasing, and their differences from $\mu_h$ are $B_h$-Lipschitz. The same properties for $Q_c$ follow from \eqref{eq:Q}.
    Finally, $x \mapsto a-\mu_h(\bar x)+\mu_h(x)-B_h(\bar x-x)$ is weakly increasing and its difference from $\mu_h$ is $B_h$-Lipschitz.
    Since the pointwise maximum preserves these common increment bounds, $f_a$ is weakly increasing and $x\mapsto\mu_h(x)-f_a(x)$ is $B_h$-Lipschitz on $[\ell,h)$.

    By construction, $f_a(x)\geq Q_\ell(x)\geq\mu_h(x)$ for $x\in[\ell,\bar x]$ and $f_a(x)\geq Q_{\bar x}(x)\geq\mu_h(x)$ for $x\in[\bar x,h)$, so group-monotonicity is satisfied.

    It remains to verify the support restriction.
    On $[\ell,\bar x]$, $f_a$ is weakly increasing and
    $f_a(\bar x)=a\leq M_2$, and hence $f_a(x)\leq M_2$.
    For $x\in[\bar x,h)$, admissibility of $b$ implies $a+V_{B_h}^{+}(\bar x,x)\le a+\sup_{x\in[\bar x, h)}V_{B_h}^{+}(\bar x,x) \le M_2$.
    To bound the other component, take any feasible counterfactual mean function $\mu_{0,\ell}^*$, whose existence follows from the maintained assumptions. For every $\bar x\leq z\leq x<h$, \eqref{eq:V} and group-monotonicity imply $\mu_{0,\ell}^*(x) \ge \mu_{0,\ell}^*(z) + V_{B_h}^{+}(z,x) \ge \mu_h(z) + V_{B_h}^{+}(z,x)$. Taking the supremum over $z\in[\bar x,x]$ gives $Q_{\bar x}(x) \le\mu_{0,\ell}^*(x) \le M_2$.
    Hence $f_a(x)\leq M_2$ on $[\bar x,h)$ as well. The lower support restriction follows from $f_a(x)\geq\mu_h(x)\geq M_1$.

    We can therefore define $\mu_{0,\ell}$ using $f_a$ over $[\ell,h)$ and extend $\mu_{0,\ell}$ and $\mu_{0,h}$ beyond $h$ continuously as in part (ii), while preserving RV-monotonicity, the smooth-bias restriction, and the support
    restriction.
    The resulting completion also satisfies group-monotonicity and attains $\mu_{0,\ell}(\bar x)=a$.
    Hence every $b\in\mathcal B_{\mathtt{MS2}}(\bar x)$ is attainable, proving sharpness.

\noindent\textit{Proof of parts (ii)$^\prime$ and (iii)$^\prime$.}
    Parts (ii)$^\prime$ and (iii)$^\prime$ of Theorem~\ref{thm:main2-OA} are analogous, with the inequalities are reversed and $V_{B_h}^-$ replacing $V_{B_h}^+$.
\end{proof}

\paragraph{Comparing Sharp and Intersection Bounds.}
This part compares the intersection bounds $\mathcal B_{\mathtt{M1}}(\bar x) \cap \mathcal B_{\mathtt S}(\bar x)$ and the sharp bounds $\mathcal B_{\mathtt{MS1}}(\bar x)$.
Since $V_{B_h}^+$ is the positive variation of a function $\mu_h(x) - B_h x$, it is instructive to rewrite this as the integral form.
To this aim, suppose additionally that $\mu_h$ is absolutely continuous.
Then, $V_{B_h}^{+}(s,t)$ is rewritten by $V_{B_h}^{+}(s,t) = \int_s^t [\mu_h^\prime(x) - B_h]_+\,dx$.
Hence, the upper endpoint of $\mathcal B_{\mathtt{MS1}}(\bar x)$ is given by
\begin{align*}
    \mu_\ell(\bar x) - \lim_{x\uparrow\ell}\mu_\ell(x) - \int_{\ell}^{\bar x} [\mu_h^\prime(x) - B_h]_+\,dx,
\end{align*}
whereas the upper endpoint of the valid intersection bound $\mathcal B_{\mathtt{M1}}(\bar x) \cap \mathcal B_{\mathtt S}(\bar x)$ in Theorem~\ref{thm:main2}(ii) is given by
\begin{align*}
    &\{\mu_\ell(\bar x) - \lim_{x\uparrow \ell}\mu_\ell(x)\}\wedge
    \{\mu_\ell(\bar x)-\mu_h(\bar x)+\mu_h(\ell)-\lim_{x\uparrow\ell}\mu_\ell(x)+B_h(\bar x-\ell)\}\\
    &=
    \mu_\ell(\bar x) - \lim_{x\uparrow \ell}\mu_\ell(x) -
    \left[\mu_h(\bar x) - \mu_h(\ell) - B_h(\bar x-\ell)\right]_+\\
    &=
    \mu_\ell(\bar x) - \lim_{x\uparrow \ell}\mu_\ell(x) - \left[
    \int_{\ell}^{\bar x} \left(\mu_h^\prime(x) - B_h\right)\,dx
    \right]_+.
\end{align*}
Since $\int_{\ell}^{\bar x} [\mu_h^\prime(x) - B_h]_+\,dx \ge [\int_{\ell}^{\bar x} (\mu_h^\prime(x) - B_h)\,dx]_+$ in general, the upper bound of $\mathcal B_{\mathtt{MS1}}(\bar x)$ is weakly tighter.
The inequality holds with equality if and only if the sign of $\mu_h^\prime(x) - B_h$ does not change sign almost everywhere on $[\ell,\bar x]$.

The relative tightness of the lower endpoints of $\mathcal B_{\mathtt{MS1}}(\bar x)$ and $\mathcal B_{\mathtt{M1}}(\bar x) \cap \mathcal B_{\mathtt S}(\bar x)$ is determined by the relative magnitudes of $L_{\mathtt{MS1}} := \sup_{x\in[\bar x,h)}\{\mu_\ell(\bar x) - \mu_h(x) + V_{B_h}^{+}(\bar x,x)\}$ and $L_{\mathtt{M1}} := \sup_{x\in[\bar x,h)}\{\mu_\ell(\bar x) - \mu_h(x)\}$, which are computed as
\begin{align*}
    L_{\mathtt{MS1}} 
    &=\mu_\ell(\bar x) - \inf_{x\in[\bar x,h)}\{\mu_h(x) - V_{B_h}^{+}(\bar x,x)\}\\
    &=\mu_\ell(\bar x) - \mu_h(\bar x) - \inf_{x\in[\bar x,h)} \int_{\bar x}^x \left\{\mu_h^\prime(z) - [\mu_h^\prime(z) - B_h]_+\right\}\,dz\\
    &=\mu_\ell(\bar x) - \mu_h(\bar x) - \inf_{x\in[\bar x,h)} \int_{\bar x}^x \min\left\{\mu_h^\prime(z), \,B_h\right\}\,dz,
\end{align*}
and
\begin{align*}
    L_{\mathtt{M1}} 
    =\mu_\ell(\bar x) - \inf_{x\in[\bar x,h)}\mu_h(x)
    =\mu_\ell(\bar x) - \mu_h(\bar x) - \inf_{x\in[\bar x,h)} \int_{\bar x}^x \mu_h^\prime(z)\,dz.
\end{align*}
Again, we can see that $L_{\mathtt{MS1}} \geq L_{\mathtt{M1}}$ in general, and hence the lower bound of $\mathcal B_{\mathtt{MS1}}(\bar x)$ is weakly tighter.
If $\mu_h^\prime(x)-B_h$ does not change sign almost everywhere on $[\bar x, h)$, then $L_{\mathtt{MS1}}=L_{\mathtt{M1}}$, so that the lower endpoints of the two identified sets coincide. In contrast to the case of the upper bound, however, this is only a sufficient condition.
We can also see that the lower bounds coincide if $\mu_h$ is weakly increasing.

\subsection{Proofs of Propositions}
\begin{proof}[Proof of Proposition \ref{prop:rv}]
    We only show the sharpness of \eqref{eq:bounds1}.
    Recall that $\tau_{\ell}(\bar{x}) = \mu_{1,\ell}(\bar{x})-\mu_{0,\ell}(\bar{x}) 
    = \mu_{\ell}(\bar{x}) - \mu_{0,\ell}(\bar{x})$.
    It now suffices to see that $[\lim_{x\uparrow \ell}\mu_{\ell}(x), M_2]$ is the sharp identified set of $\mu_{0,\ell}(\bar x)$.
    The validity of the identified set is immediate by observing that $\lim_{x\uparrow \ell}\mu_{\ell}(x) = \mu_{0,\ell}(\ell) \le \mu_{0,\ell}(\bar x) \le M_2$.
    To see the sharpness, take any point $a \in [\lim_{x\uparrow \ell}\mu_{\ell}(x), M_2]$. Then we can define 
    \begin{align*}
        \mu_{0,\ell}(x) = \begin{cases}
            \mu_{\ell}(x) & \text{ if } x \in (-\infty, \ell) \cap \mathcal X,\\
            \dfrac{a - \lim_{x\uparrow \ell}\mu_{\ell}(x)}{\bar{x} - \ell} (x - \ell) + \lim_{x\uparrow \ell}\mu_{\ell}(x) & \text{ if } x \in [\ell, \bar x],\\
            a & \text{ if } x \in (\bar x, \infty) \cap \mathcal X.
        \end{cases}
    \end{align*}
    Then, this $\mu_{0,\ell}$ satisfies Assumptions \ref{assumption:continuity} and \ref{assumption:x-monotonicity}(i).
    Furthermore, it holds that $\mu_{0,\ell}(\bar x) = a$, showing that $a \in [\lim_{x\uparrow \ell}\mu_{\ell}(x), M_2]$ is attainable. Since $a$ is arbitrary, the sharpness follows.
\end{proof}

\begin{proof}[Proof of Proposition \ref{prop:group}]
    The proof is similar to that of Proposition \ref{prop:rv}.
\end{proof}

\begin{proof}[Proof of Proposition \ref{prop:smooth-single}]
    We prove part (i) and the first half of (ii).
    We first establish validity of the stated bounds.
    In case (i), Assumption \ref{assumption:lipchitz-l} and continuity imply $|\mu_{0,\ell}(\bar x)-\lim_{x\uparrow\ell}\mu_\ell(x)| = |\mu_{0,\ell}(\bar x)-\mu_{0,\ell}(\ell)| \leq C_\ell(\bar x-\ell)$.
    Together with the support restriction, it follows that
    \begin{align*}
        M_1
        \vee
        \left\{
        \lim_{x\uparrow\ell}\mu_\ell(x)
        -
        C_\ell(\bar x-\ell)
        \right\}
        \leq \mu_{0,\ell}(\bar x)
        \leq
        M_2
        \wedge
        \left\{
        \lim_{x\uparrow\ell}\mu_\ell(x)
        +
        C_\ell(\bar x-\ell)
        \right\}.
    \end{align*}
    Since $\tau_\ell(\bar x) = \mu_\ell(\bar x) - \mu_{0,\ell}(\bar x)$, this yields the bounds in part (i).
    In case (ii), RV-monotonicity additionally implies $\mu_{0,\ell}(\bar x) \ge \mu_{0,\ell}(\ell) = \lim_{x\uparrow\ell}\mu_\ell(x)$.
    Combining this restriction with the Lipschitz and support restrictions gives
    \begin{align*}
        \lim_{x\uparrow\ell}\mu_\ell(x)
        \leq
        \mu_{0,\ell}(\bar x)
        \leq
        M_2
        \wedge
        \left\{
        \lim_{x\uparrow\ell}\mu_\ell(x)
        +
        C_\ell(\bar x-\ell)
        \right\},
    \end{align*}
    which is equivalent to the bounds stated in part (ii).

    It remains only to establish sharpness.
    Fix any $b$ in the identified set stated in the corresponding part of the proposition, and let $a:=\mu_\ell(\bar x)-b$.
    Define a completion of the unobserved conditional mean function by
    \begin{align*}
        \mu_{0,\ell}(x)
        =
        \begin{cases}
            \mu_\ell(x), & \text{if } x\in(-\infty,\ell)\cap\mathcal X,\\
            \dfrac{a-\lim_{z\uparrow\ell}\mu_\ell(z)}{\bar x-\ell}(x-\ell)
            +
            \lim_{z\uparrow\ell}\mu_\ell(z), & \text{if } x\in[\ell,\bar x],\\
            a, & \text{if } x\in(\bar x,\infty)\cap\mathcal X.
        \end{cases}
    \end{align*}
    By construction, the function is continuous and satisfies $\mu_{0,\ell}(\bar x)=a$. Moreover, its absolute slope on $[\ell,\bar x]$ is no greater than $C_\ell$ by admissibility of $a$, while the function is constant for $x>\bar x$. Hence, $\mu_{0,\ell}\in\mathcal L(C_\ell)$.
    Since both endpoint values $\lim_{z\uparrow\ell}\mu_\ell(z)$ and $a$ belong to $[M_1,M_2]$, the constructed function also takes values in $[M_1,M_2]$.

    In case (ii), admissibility of $a$ further implies $a\geq\lim_{z\uparrow\ell}\mu_\ell(z)$. Thus the constructed function is weakly increasing on $[\ell,\infty)\cap\mathcal X$, and therefore satisfies Assumption \ref{assumption:x-monotonicity}(i).
    
    The construction changes $\mu_{0,\ell}(x)$ only for $x\geq\ell$ and hence does not alter the distribution of observed outcomes. On this unobserved region, one can choose a conditional distribution of $Y_i(0)$ having the constructed mean and support contained in $[M_1,M_2]$, while leaving all observed conditional distributions unchanged.
    Therefore every admissible value $a$ is attainable, proving sharpness of the identified sets. 
\end{proof}

\section{Multi-Cutoff Designs}\label{sec:multi}

This section provides identification results for the multiple-cutoff design. We first extend Theorem~\ref{thm:main1} to this setting and then discuss identification under alternative assumptions. We subsequently establish valid identified sets that combine the monotonicity restrictions with smoothness assumptions.

\subsection{Setup}
We begin by introducing the setup.
Let $Y_i \in \mathbb R$ denote the outcome variable, $X_i \in \mathcal X \subseteq \mathbb R$ the running variable with support $\mathcal X$, and $D_i\in\{0,1\}$ be a treatment indicator, which takes one if $i$ is treated and zero otherwise. 
Treatment assignment follows the rule $D_i = \mathbf{1}\{X_i \geq C_i\}$, where $C_i \in \mathcal C$ denotes the cutoff position, and
\begin{align*}
    \mathcal C = \{c_{-L},\ldots, c_0, \ldots, c_{U}\},\quad
    c_{-L} < \cdots < c_{U},
\end{align*}
denotes the collection of threshold points. 
Let $Y_i(d)$ denote the potential outcome under treatment status $D_i=d\in\{0,1\}$. 
Define the (potential) conditional expectations and treatment effect functions as
\begin{align*}
    \mu_{d,c}(x) &\coloneqq \E{Y_i(d) | X_i = x, C_i = c},\\
    \tau_c(x) &\coloneqq \E{Y_i(1) - Y_i(0) | X_i = x, C_i = c} = \mu_{1,c}(x)-\mu_{0,c}(x).
\end{align*}
Write the observed regression function by
\begin{align*}
    \mu_{c}(x) \coloneqq \E{Y_i | X_i = x, C_i = c}.
\end{align*}
We maintain the following condition which assumes the continuity of the potential mean functions:
\begin{assumption}[Continuity]\label{assumption:continuity-OA}
    $\mu_{d,c}(x)$ is continuous in $x \in \mathcal X$ for  $d\in\{0,1\}$ and $c\in\mathcal C$.
\end{assumption}

\subsection{Identification Results under Monotonicity Assumptions}

We first study the identification power of RV- and group-monotonicity conditions. 
Without loss of generality, suppose that group $c_0\in\mathcal C = \{c_{-L},\ldots, c_0, \ldots, c_{U}\}$ is our target group.

\begin{assumption}[Running-Variable (RV) Monotonicity]\label{assumption:x-monotonicity-OA}
    Either of the following conditions holds.
    \begin{description}
        \item[(i)] For any $c \in \mathcal C$, $\mu_{0,c}(x)$ is weakly increasing in $x \in [c,\infty)\cap\mathcal X$.
        \item[(ii)] For any $c \in \mathcal C$, $\mu_{0,c}(x)$ is weakly decreasing in $x \in [c,\infty)\cap\mathcal X$.
    \end{description}
\end{assumption}

\begin{assumption}[Group Monotonicity]\label{assumption:c-monotonicity-OA}
    Either of the following conditions hold.
    \begin{description}
        \item[(i)] For $c,c^\prime \in \mathcal C$, $c \leq c^\prime \Rightarrow \mu_{0,c}(x) \leq \mu_{0,c^\prime}(x)$ for $x\in[c, c^\prime]$.
        \item[(ii)] For $c,c^\prime \in \mathcal C$, $c \leq c^\prime \Rightarrow \mu_{0,c}(x) \geq \mu_{0,c^\prime}(x)$ for $x\in[c, c^\prime]$.
    \end{description}
\end{assumption}

\begin{prop}\label{prop:three}
    Let $c_0 \le c_{k^*} < \bar x \le c_{k^* + 1} \le c_U$. Suppose that Assumption~\ref{assumption:continuity-OA} holds true.
    For each cutoff group $c\in\mathcal C$, define the observed untreated extension
    \begin{align*}
        \mu_c^{-}(x)
        :=
        \begin{cases}
            \mu_c(x), & x<c,\\
            \lim_{z\uparrow c}\mu_c(z), & x=c,
        \end{cases}
    \end{align*}
    for $x\le c$.

    \begin{description}
        \item[(i)] If Assumption~\ref{assumption:x-monotonicity-OA}(i) holds, then the sharp identified set of $\tau_{c_0}(\bar x)$ is given by 
        \begin{align*}
            \left[
            \mu_{c_0}(\bar x) - M_2,\, \mu_{c_0}(\bar x) - \lim_{x\uparrow c_0}\mu_{c_0}(x)
            \right],
        \end{align*}
        provided that $Y_i(0)\in [M_1, M_2]$.
        If Assumption~\ref{assumption:x-monotonicity-OA}(ii) holds instead, the sharp identified set of $\tau_{c_0}(\bar x)$ is given by 
        \begin{align*}
            \left[
            \mu_{c_0}(\bar x) - \lim_{x\uparrow c_0}\mu_{c_0}(x),\,
            \mu_{c_0}(\bar x) - M_1
            \right],
        \end{align*}
        provided that $Y_i(0)\in [M_1, M_2]$.

        \item[(ii)] If Assumption~\ref{assumption:c-monotonicity-OA}(i) holds, then the sharp identified set of $\tau_{c_0}(\bar x)$ is given by 
        \begin{align*}
            \left[
            \mu_{c_0}(\bar x) - \min_{j\in\{k^*+1,\ldots,U\}} \mu_{c_j}^{-}(\bar x),\,
            \mu_{c_0}(\bar x) - M_1
            \right],
        \end{align*}
        provided that $Y_i(0)\in [M_1, M_2]$.
        If Assumption~\ref{assumption:c-monotonicity-OA}(ii) holds instead, the sharp identified set of $\tau_{c_0}(\bar x)$ is given by 
        \begin{align*}
            \left[
            \mu_{c_0}(\bar x)-M_2,
            \,
            \mu_{c_0}(\bar x)
            -
            \max_{j\in \{k^*+1,\ldots,U\}}\mu_{c_j}^{-}(\bar x)
            \right],
        \end{align*}
        provided that $Y_i(0)\in [M_1, M_2]$.

        \item[(iii)] If Assumptions \ref{assumption:x-monotonicity-OA}(i) and \ref{assumption:c-monotonicity-OA}(i) holds, then the sharp identified set of $\tau_{c_0}(\bar x)$ is given by 
        \begin{align*}
            \left[
            \mu_{c_0}(\bar x) - \min_{j\in\{k^*+1,\ldots,U\}} \inf_{x \in[\bar x, c_j]} \mu_{c_j}^{-}(x),\, 
            \mu_{c_0}(\bar x) - \lim_{x\uparrow c_0}\mu_{c_0}(x)
            \right].
        \end{align*}
        If Assumptions \ref{assumption:x-monotonicity-OA}(ii) and \ref{assumption:c-monotonicity-OA}(ii) holds instead, then the sharp identified set of $\tau_{c_0}(\bar x)$ is given by 
        \begin{align*}
            \left[
            \mu_{c_0}(\bar x) - \lim_{x\uparrow c_0}\mu_{c_0}(x),\,
            \mu_{c_0}(\bar x) - \max_{j\in \{k^*+1,\ldots,U\}}\sup_{x\in[\bar x,c_j]}\mu_{c_j}^{-}(x)
            \right].
        \end{align*}

        \item[(iv)] If Assumptions \ref{assumption:x-monotonicity-OA}(i) and \ref{assumption:c-monotonicity-OA}(ii) holds, then the sharp identified set of $\tau_{c_0}(\bar x)$ is given by 
        \begin{align*}
            \bigg[
            \mu_{c_0}(\bar x)-M_2,\,
            &\left(
            \mu_{c_0}(\bar x)
            -
            \max_{j\in\{1,\ldots, k^*\}}\sup_{x\in[c_0, c_j]}\mu_{c_j}^{-}(x)
            \right)\\
            &\quad\wedge
            \left(
            \mu_{c_0}(\bar x)
            -
            \max_{j\in \{k^*+1,\ldots,U\}}\sup_{x\in[c_0, \bar x]}\mu_{c_j}^{-}(x)
            \right)
            \wedge
            \left(
            \mu_{c_0}(\bar x)
            -
            \lim_{x\uparrow c_0}\mu_{c_0}(x)
            \right)
            \bigg],
        \end{align*}
        provided that $Y_i(0)\in [M_1, M_2]$.
        If Assumptions \ref{assumption:x-monotonicity-OA}(ii) and \ref{assumption:c-monotonicity-OA}(i) holds instead, then the sharp identified set of $\tau_{c_0}(\bar x)$ is given by 
        \begin{align*}
            \bigg[
            &\left(
            \mu_{c_0}(\bar x)
            -
            \min_{j\in\{1,\ldots,k^*\}}\inf_{x\in[c_0, c_j]}\mu_{c_j}^{-}(x)
            \right)\\
            &\quad\vee
            \left(
            \mu_{c_0}(\bar x)
            -
            \min_{j\in \{k^*+1,\ldots,U\}}\inf_{x\in[c_0, \bar x]}\mu_{c_j}^{-}(x)
            \right)
            \vee
            \left(
            \mu_{c_0}(\bar x)
            -
            \lim_{x\uparrow c_0}\mu_{c_0}(x)
            \right),
            \,
            \mu_{c_0}(\bar x)-M_1
            \bigg],
        \end{align*}
        provided that $Y_i(0)\in [M_1, M_2]$.
    \end{description}
\end{prop}

\begin{remark}
    As may be clear from part (ii), relative to using any single eligible comparison group, combining information from multiple higher-cutoff groups weakly tightens the bounds.
\end{remark}

\begin{remark}\label{rem:strong}
    As discussed in Theorem~\ref{thm:main1}(iii), if $\mu_{0,c}$ satisfies the same monotonicity restriction with respect to the running variable over $\mathcal X$ for every $c\in\mathcal C$, the bounds in Proposition~\ref{prop:three}(iii)--(iv) can be simplified.
    For example, consider the first case in part (iii). If $\mu_{0,c}$ is weakly increasing over $\mathcal X$ for every $c\in\mathcal C$, the sharp identified set can be rewritten as
    \begin{align*}
        \left[
        \mu_{c_0}(\bar x) - \min_{j\in\{k^*+1,\ldots,U\}} \mu_{c_j}^{-}(\bar x),\, 
        \mu_{c_0}(\bar x) - \lim_{x\uparrow c_0}\mu_{c_0}(x)
        \right].
    \end{align*}
    If, in addition, the group monotonicity ordering holds over the entire support $\mathcal X$, the bounds further simplify to
    \begin{align*}
        \left[
        \mu_{c_0}(\bar x) - \mu_{c_{k^*+1}}^{-}(\bar x),\, 
        \mu_{c_0}(\bar x) - \lim_{x\uparrow c_0}\mu_{c_0}(x)
        \right].
    \end{align*}
\end{remark}

A graphical example is useful for illustrating which groups' treatment effects can be identified and over which regions of the running variable.
\begin{figure}[t]
    \centering
    \begin{subfigure}[b]{0.49\textwidth}
        \centering
        \includegraphics[width=\linewidth]{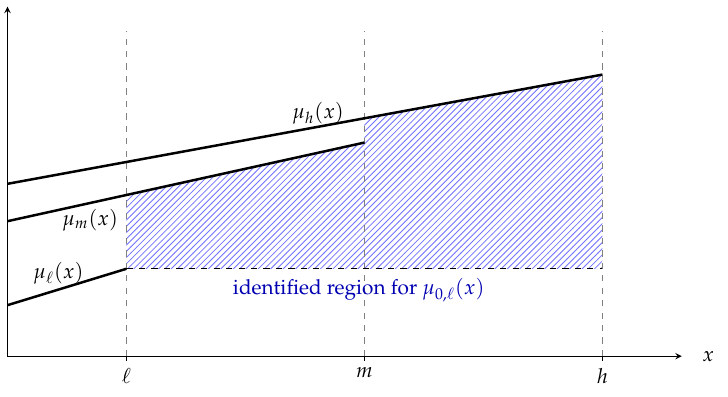}
        \caption{Bounds on $\mu_{0,\ell}(x)$}
        \label{fig:tikz6}
    \end{subfigure}
    \hfill
    \begin{subfigure}[b]{0.49\textwidth}
        \centering
        \includegraphics[width=\linewidth]{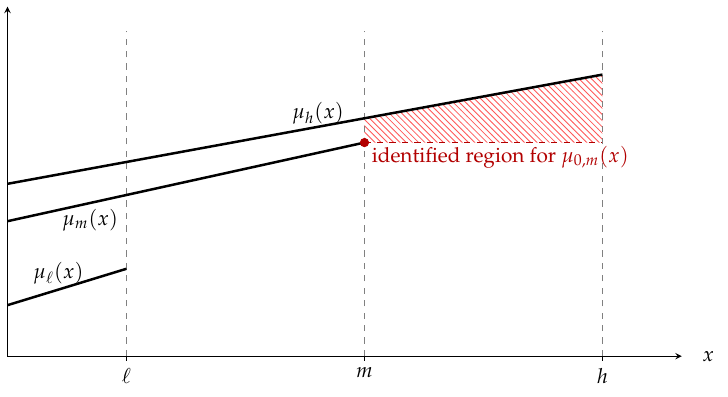}
        \caption{Bounds on $\mu_{0,m}(x)$ and $\mu_{0,h}(x)$}
        \label{fig:tikz7}
    \end{subfigure}
    \caption{Bounds under Monotonicity Assumptions with Multiple Cutoffs}
    \label{fig:three}

\end{figure}
Figure~\ref{fig:three} provides an illustration of a three-cutoff design with cutoffs $\ell < m < h$.

Suppose Assumptions~\ref{assumption:x-monotonicity-OA}(i) and \ref{assumption:c-monotonicity-OA}(i) hold. We further impose the stronger conditions in Remark~\ref{rem:strong} to simplify the exposition.
For group $\ell$, if the target score value satisfies $\bar x\in(\ell,m]$, the counterfactual conditional mean is bounded as $[\lim_{x\uparrow \ell}\mu_{\ell}(x), \mu_{m}^{-}(\bar x)]$.
If instead $\bar x\in(m,h]$, then $\mu_{0,\ell}(\bar x) \in [\lim_{x\uparrow \ell}\mu_{\ell}(x), \mu_{h}^{-}(\bar x)]$.
For group $m$, we obtain analogous bounds $\mu_{0,m}(\bar x) \in [\lim_{x\uparrow m}\mu_{m}(x), \mu_{h}^{-}(\bar x)]$ but only for $\bar x\in(m,h]$.

For a given target group, once the evaluation point lies above all available higher cutoffs, no untreated higher-cutoff comparison group remains available at that point. The identification problem then reduces to the single-cutoff setting and generally yields only one-sided bounds.

\subsection{Alternative Group Restrictions}\label{sec:alter}
Let $c_0$ denote the target group, as above.
In some applications, imposing group monotonicity across all cutoff groups such that $c \ge c_0$ may be overly restrictive. We therefore note two alternatives.
First, suppose that there exists a subset $\mathcal C_+ \subseteq \{c\in\mathcal C: c \ge \bar x\}$ such that
\begin{align*}
    \mu_{0,c_0}(\bar x)
    \leq
    \mu_{0,c}(\bar x)
    \qquad
    \text{for every } c\in\mathcal C_+.
\end{align*}
This restriction does not require an ordering among all cutoff groups, but only requires the target group to lie below a prespecified collection of comparison groups.
For example, $c_0$ may represent a relatively disadvantaged rural region, whereas $\mathcal C_+$ consists of affluent metropolitan regions.
In such a setting, it may be plausible to posit that untreated potential outcomes in the target region are weakly lower than those in each of these urban regions, even if no meaningful ordering is available for the remaining groups.
Then, $\mu_{0,c_0}(\bar x) \le \min_{c \in \mathcal C_+}\mu_{c}^{-}(\bar x)$, and hence $\tau_{c_0}(\bar x) \ge \mu_{c_0}(\bar x) - \min_{c \in \mathcal C_+}\mu_{c}^{-}(\bar x)$.
If the RV-monotonicity is assumed as well, a simple valid identified set is 
\begin{align*}
    \tau_{c_0}(\bar x) \in \left[
    \mu_{c_0}(\bar x) - \min_{c \in \mathcal C_+}\mu_{c}^{-}(\bar x),\,
    \mu_{c_0}(\bar x) - \lim_{x\uparrow c_0}\mu_{c_0}(x)
    \right].
\end{align*}

A still weaker alternative is to assume only that
\begin{align}
    \mu_{0,c_0}(\bar x)
    \leq
    \mu_{0,c}(\bar x)
    \qquad
    \text{for at least one } c\in\mathcal C_+,\label{eq:rem:union}
\end{align}
without specifying which comparison group satisfies the inequality.
Such a restriction may be appropriate when institutional or socioeconomic information suggests that the target group is not the most advantaged group within
$\{c_0\}\cup\mathcal C_+$, while the relative ordering of the comparison groups themselves is ambiguous.
For example, $c_0$ may represent a low-income rural region, whereas $\mathcal C_+$ consists of metropolitan regions that differ in income, industrial composition, and demographic characteristics.
It may then be credible that at least one of these metropolitan regions has a higher untreated potential outcome than the target region, even though it would be difficult to maintain this ordering for every region in $\mathcal C_+$.
Under this weaker restriction, $\mu_{0,c_0}(\bar x) \le \max_{c \in \mathcal C_+}\mu_{c}^{-}(\bar x)$, which yields $\tau_{c_0}(\bar x) \ge \mu_{c_0}(\bar x) - \max_{c \in \mathcal C_+}\mu_{c}^{-}(\bar x)$.
If the RV-monotonicity is assumed as well, we obtain valid bounds
\begin{align*}
    \tau_{c_0}(\bar x) \in \left[
    \mu_{c_0}(\bar x) - \max_{c \in \mathcal C_+}\mu_{c}^{-}(\bar x),\,
    \mu_{c_0}(\bar x) - \lim_{x\uparrow c_0}\mu_{c_0}(x)
    \right].
\end{align*}

There may be other plausible restrictions. We leave a detailed investigation of this interesting issue for future research.

\subsection{Combining Smoothness Assumptions}

As discussed in the main article, smoothness restrictions can provide additional identifying power. Here, rather than pursuing sharpness, we construct valid, though potentially non-sharp, bounds by intersecting the identified sets implied by different restrictions.

Let $c_0 \in \mathcal C$ denote the target group.
For each $c>c_0$, define the bias function $\delta_c(x):=\mu_{0,c}(x)-\mu_{0,c_0}(x)$.
We impose the following smooth-bias assumption introduced in the main article.

\begin{assumption}[Smooth Bias]\label{assumption:lipchitz-bias-OA}
    For each $c\in\mathcal C$ such that $c>c_0$, $\delta_c \in \mathcal{L}(B_c)$, where
    \begin{align*}
        \mathcal{L}(C) = \left\{
        \mu : \left|\mu(x_1) - \mu(x_2)\right| \leq C|x_1 - x_2| \text{ for all } x_1, x_2 \in [c_0, \infty) \cap \mathcal X
        \right\}
    \end{align*}
    and $B_c\geq 0$ is a known constant.
\end{assumption}

For a target value $\bar x\in(c_{k^*},c_{k^*+1}]$, each cutoff group $c_k$ with $k\geq k^*+1$ provides a smooth-bias-based identified set for $\tau_{c_0}(\bar x)$. Intersecting these sets yields the valid identified region
\begin{align*}
    \bigcap_{k\geq k^*+1} \mathcal B_{\mathtt S}^{c_k}(\bar x),
\end{align*}
where
\begin{align*}
    \mathcal B_{\mathtt S}^{c}(\bar x)
    :=
    \left[
    \tau_{c_0}^{\mathtt{CB},c}(\bar x)-B_c(\bar x-c_0),\,
    \tau_{c_0}^{\mathtt{CB},c}(\bar x)+B_c(\bar x-c_0)
    \right],
\end{align*}
and $\tau_{c_0}^{\mathtt{CB},c}(\bar x) := \mu_{c_0}(\bar x)-\mu_c^{-}(\bar x) + \mu_c(c_0) - \lim_{x\uparrow c_0}\mu_{c_0}(x)$.

A valid identified set under both the monotonicity and smooth-bias restrictions is obtained by intersecting
$\bigcap_{k\geq k^*+1}\mathcal B_{\mathtt S}^{c_k}(\bar x)$
with the monotonicity-based bounds provided in Proposition~\ref{prop:three}.

\subsection{Outline of the Inference Procedure}

The same inference procedures discussed in the main text may be applied here.
For example, the bounds in Proposition~\ref{prop:three}(i)--(ii) admit polyhedral representations, whereas those in Proposition~\ref{prop:three}(iii)--(iv) can be enclosed by valid outer bounds that are polyhedral, following the same construction as in Example~\ref{ex:outer-approx}. Hence, the moment-inequality-based inference of \cite{Andrews_etal:2023} can be employed.

Related to the restriction \eqref{eq:rem:union} in Section~\ref{sec:alter}, the identified set may in some cases be represented as a union, rather than an intersection, of identified sets. For example,
\begin{align*}
    \tau_{c_0}(\bar x)
    &\in
    \left[
    \mu_{c_0}(\bar x)-\max_{c\in\mathcal C_+}\mu_c^{-}(\bar x),\,
    \mu_{c_0}(\bar x) - \lim_{x\uparrow c_0}\mu_{c_0}(x)
    \right]\\
    &=
    \bigcup_{c\in\mathcal C_+}
    \left[
    \mu_{c_0}(\bar x)-\mu_c^{-}(\bar x),\,\mu_{c_0}(\bar x) - \lim_{x\uparrow c_0}\mu_{c_0}(x)
    \right].
\end{align*}
In such a case, a valid confidence set can be constructed by taking the union of the confidence sets corresponding to the individual sets appearing on the right-hand side.
More generally, suppose that an identified set can be represented as a finite union of polyhedrally represented identified sets. We can apply the moment-inequality-based inference procedure to each component separately and then take the union of the resulting confidence sets. The resulting set provides a valid confidence set for the original union-identified set (\citealp{Rambachan_Roth:2023}).
Thus, the same moment-inequality-based approach still applies.

\subsection{Proofs for Appendix \ref{sec:multi}}
\begin{proof}[Proof of Proposition \ref{prop:three}]
    \textit{Proof of part (i)}. With no cross-group restrictions imposed, the only source of identifying power comes from the RV monotonicity of $\mu_{0,c_0}$. Hence, the sharp identified set coincides with that in Proposition \ref{prop:rv}.

    \noindent \textit{Proof of part (ii)}. 
    For $\bar x \in (c_{k^*}, c_{k^*+1}]$, the available restrictions on $\mu_{0,c_0}(\bar x)$ are $\mu_{0,c_0}(\bar x) \le \mu_{0,c_{k^*+1}}(\bar x) = \mu_{c_{k^*+1}}^{-}(\bar x), \ldots, \mu_{0,c_0}(\bar x) \le \mu_{0,c_{U}}(\bar x) = \mu_{c_{U}}^{-}(\bar x)$. Together with the support restriction, the bounds are valid. 

    To establish sharpness, suppose first that Assumption~\ref{assumption:c-monotonicity-OA}(i) holds.
    Fix any $b$ in the stated identified set and let $a:=\mu_{c_0}(\bar x)-b$. Then $M_1 \le a \le \min_{j\in\{k^*+1,\ldots,U\}} \mu_{c_j}^{-}(\bar x)$.
    Take any feasible collection of counterfactual mean functions $\{\mu_{0,c}^*\}_{c\in\mathcal C}$ satisfying the maintained assumptions. Choose a sufficiently small neighborhood $\mathcal I$ of $\bar x$ such that $\mathcal I\subset(c_{k^*},\infty)$ and $\mathcal I$ contains no cutoff other than possibly $\bar x=c_{k^*+1}$. Let $\psi:\mathcal I\to[0,1]$ be continuous, equal to one at $\bar x$, and equal to zero at the boundary of $\mathcal I$.
    
    For $x\in \mathcal I$, define the set of comparison groups whose group-ordering restriction applies at $x$ by $\mathcal J(x) := \{j\in\{1,\ldots,U\}:x\leq c_j\}$.
    Now set
    \begin{align*}
        \widetilde\mu_{0,c_0}(x) :=
        \min\left\{(1-\psi(x))\mu_{0,c_0}^*(x)+\psi(x)a,\, 
        \min_{j\in\mathcal J(x)} \mu_{0,c_j}^*(x)\right\},
        \qquad x\in \mathcal I,
    \end{align*}
    where $\min_{j\in\mathcal J(x)} \mu_{0,c_j}^*(x)=M_2$ when $\mathcal J(x)$ is empty, and let $\widetilde\mu_{0,c_0}(x) := \mu_{0,c_0}^*(x)$ for $x\notin \mathcal I$.
    Since the original counterfactual mean is feasible, $\mu_{0,c_0}^*(x)\leq \min_{j\in\mathcal J(x)} \mu_{0,c_j}^*(x)$ whenever the comparison set is nonempty. Hence the modified function agrees with $\mu_{0,c_0}^*$ at the boundary of $\mathcal I$.
    Moreover, $\min_{j\in\{k^*+1,\ldots,U\}} \mu_{c_j}^{-}(\bar x) \ge a$ and therefore $\widetilde\mu_{0,c_0}(\bar x)=a$.
    The construction is continuous:
    In particular, if $\bar x=c_{k^*+1}$, the group $c_{k^*+1}$ drops from $\mathcal J(x)$ immediately to the right of $\bar x$, so that $\min_{j\in\mathcal J(x)} \mu_{0,c_j}^*(x)$ may jump upward there; nevertheless, since $a\leq \min_{j\in\mathcal J(x)} \mu_{0,c_j}^*(\bar x)$ and the first argument of the minimum converges to $a$ from both sides, $\widetilde\mu_{0,c_0}$ remains continuous at $\bar x$.
    
    By construction, $\widetilde\mu_{0,c_0}(x) \le \mu_{0,c_j}^*(x)$ whenever $x\in[c_0,c_j]$, so all group-monotonicity restrictions involving the target group are preserved. All restrictions involving only the remaining groups are unchanged. Furthermore, both arguments of the minimum take values in $[M_1,M_2]$, so the support restriction is also preserved.
    The modification affects only the unobserved untreated potential-outcome regression of the target group and therefore leaves the distribution of observables unchanged.
    On the modified region, one can choose a conditional distribution of $Y_i(0)$ with mean $\widetilde\mu_{0,c_0}(x)$ and support contained in $[M_1,M_2]$.
    Thus every $a$ is attainable, proving sharpness.
    Under Assumption~\ref{assumption:c-monotonicity-OA}(ii), the argument
    is symmetric.

    \noindent \textit{Proof of part (iii)}. Suppose first that Assumptions~\ref{assumption:x-monotonicity-OA}(i) and \ref{assumption:c-monotonicity-OA}(i) hold.
    RV monotonicity implies $\mu_{0,c_0}(\bar x) \ge \mu_{0,c_0}(c_0) = \lim_{x\uparrow c_0}\mu_{c_0}(x)$.
    Moreover, for every $j\in\{k^*+1,\ldots,U\}$ and every $x\in[\bar x,c_j]$,
    \begin{align*}
        \mu_{0,c_0}(\bar x)
        \le \mu_{0,c_0}(x)
        \le \mu_{0,c_j}(x)
        = \mu_{c_j}^{-}(x),
    \end{align*}
    where the first inequality follows from RV monotonicity, and the second from group monotonicity.
    Hence, we have that
    \begin{align*}
        \mu_{0,c_0}(\bar x)
        \le
        \min_{j\in\{k^*+1,\ldots,U\}}\inf_{x\in[\bar x,c_j]} \mu_{c_j}^{-}(x).
    \end{align*}
    
    To establish sharpness, fix any admissible value $a$ satisfying $\lim_{x\uparrow c_0}\mu_{c_0}(x) \le a \le \min_{j\in\{k^*+1,\ldots,U\}}\inf_{x\in[\bar x,c_j]}\mu_{c_j}^{-}(x)$.
    Note that $\min_{j\in\{k^*+1,\ldots,U\}} \mu_{c_j}^{-}(x)$ is continuous on $[c_{k^*},\bar x]$, since each $\mu_{c_j}^{-}$ is continuous on $[c_{k^*},\bar x]$.
    Moreover, compatibility of the maintained assumptions implies $\lim_{x\uparrow c_0}\mu_{c_0}(x) \le \min_{j\in\{k^*+1,\ldots,U\}} \mu_{c_j}^{-}(x)$ for every $x\in[c_{k^*},\bar x]$.
    Define
    \begin{align*}
        f_a(x)
        :=
        \min\left\{\lim_{x\uparrow c_0}\mu_{c_0}(x) + \frac{a-\lim_{x\uparrow c_0}\mu_{c_0}(x)}{\bar x-c_{k^*}}(x-c_{k^*}),\,
        \inf_{z\in[x,\bar x]}\min_{j\in\{k^*+1,\ldots,U\}} \mu_{c_j}^{-}(z)\right\}
    \end{align*}
    for $x\in[c_{k^*},\bar x]$.
    Both functions inside the minimum are continuous and weakly increasing, and hence so is $f_a$. The preceding inequalities and admissibility of $a$ imply $f_a(c_{k^*})=\lim_{x\uparrow c_0}\mu_{c_0}(x)$ and $f_a(\bar x)=a$.
    Furthermore, by construction, $f_a(x) \le \min_{j\in\{k^*+1,\ldots,U\}} \mu_{c_j}^{-}(x) \le \mu_{c_j}^{-}(x)$ for every $j\in\{k^*+1,\ldots,U\}$ and $x\in[c_{k^*},\bar x]$.
    
    Now define the counterfactual mean by
    \begin{align*}
        \mu_{0,c_0}(x)
        :=
        \begin{cases}
            \mu_{c_0}(x), & x<c_0,\\
            \lim_{x\uparrow c_0}\mu_{c_0}(x), & x\in[c_0,c_{k^*}],\\
            f_a(x), & x\in[c_{k^*},\bar x],\\
            a, & x>\bar x.
        \end{cases}
    \end{align*}
    This completion is continuous and weakly increasing.
    For every $j\in\{1,\ldots,k^*\}$ and $x\in[c_0,c_j]$, compatibility of the maintained assumptions gives $\mu_{0,c_0}(x)=\lim_{x\uparrow c_0}\mu_{c_0}(x) = \mu_{0,c_0}^*(c_0) \le \mu_{0,c_0}^*(x) \le \mu_{0,c_j}^*(x) = \mu_{c_j}^{-}(x)$.
    For $j\in\{k^*+1,\ldots,U\}$, the same argument gives $\mu_{0,c_0}(x)=\lim_{x\uparrow c_0}\mu_{c_0}(x) \le \mu_{c_j}^{-}(x)$ on $[c_0,c_{k^*}]$, and the construction above guarantees the same group ordering on $[c_0,\bar x]$, while admissibility of $a$ implies $a \le \mu_{c_j}^{-}(x)$ for every $x\in[\bar x,c_j]$. 
    Hence all group-monotonicity restrictions are satisfied.
    The construction modifies only the unobserved untreated potential-outcome regression of the target group and therefore leaves the distribution of observables unchanged. Thus every admissible $a$ is attainable, proving sharpness.
    The case under Assumptions~\ref{assumption:x-monotonicity-OA}(ii) and \ref{assumption:c-monotonicity-OA}(ii) is analogous.

    \noindent \textit{Proof of part (iv)}. Suppose first that Assumptions~\ref{assumption:x-monotonicity-OA}(i) and \ref{assumption:c-monotonicity-OA}(ii) hold.
    RV monotonicity gives $\mu_{0,c_0}(\bar x) \ge \lim_{x\uparrow c_0}\mu_{c_0}(x)$.
    For a comparison group with $j\in\{1,\ldots,k^*\}$, we have $c_j<\bar x$. Hence, for every $x\in[c_0,c_j]$,
    \begin{align*}
        \mu_{0,c_0}(\bar x)
        \ge \mu_{0,c_0}(x)
        \ge \mu_{0,c_j}(x)
        = \mu_{c_j}^{-}(x).
    \end{align*}
    It then follows that
    \begin{align*}
        \mu_{0,c_0}(\bar x)
        \ge \max_{j\in\{1,\ldots,k^*\}}\sup_{x\in[c_0,c_j]} \mu_{c_j}^{-}(x).
    \end{align*}
    For a comparison group with $j\in\{k^*+1,\ldots,U\}$, the same argument applies over $x\in[c_0,\bar x]$, yielding
    \begin{align*}
        \mu_{0,c_0}(\bar x)
        \ge \max_{j\in\{k^*+1,\ldots,U\}} \sup_{x\in[c_0,\bar x]} \mu_{c_j}^{-}(x).
    \end{align*}
    Combining these inequalities with the support restriction $\mu_{0,c_0}(\bar x)\leq M_2$ gives the stated bounds.
    Sharpness follows from the same argument as in part (iii).
\end{proof}

\section{Fuzzy RD Designs}\label{sec:fuzzy}

\subsection{Setup and Identification Results}

Following \citet{Cattaneo_etal:2021JASA_extrapolating}, we extend our framework to fuzzy RD designs with one-sided noncompliance: individuals with $X_i \geq C_i$ are not necessarily treated, whereas those with $X_i<C_i$ are never treated.
This setting is empirically relevant and arises in many applications in which treatment requires active participation or enrollment.
For example, individuals may need to attend a job-training program or visit a hospital to receive a medical intervention \citep[Chapter 23]{imbens_rubin:2015causal}.
In educational RD settings, students who fall below an eligibility cutoff cannot receive a financial-aid offer, whereas those who qualify may still decline it \citep{Londono-Velez_etal:2020aejep}---a structure consistent with one-sided fuzzy RD designs.
If the one-sided noncompliance assumption fails, alternative identification strategies are required; we leave their development for future research.

Suppose that group $c_0\in\mathcal C=\{c_{-L},\ldots,c_0,\ldots,c_U\}$ is the target group.
We now define the treatment indicator as $D_i=D_i(X_i,C_i)$, with potential treatment status $D_i(x,c)\in\{0,1\}$, and impose the following assumptions:

\begin{assumption}[One-Sided Compliance Fuzzy RD]\label{assumption:fuzzy}
The following conditions hold:
\begin{description}
    \item[(i)] $\E{Y_i(0)\mid X_i=x,C_i=c}$ is continuous in $x\in\mathcal X$ for all $c\in\mathcal C$;
    \item[(ii)] $\E{Y_i\mid X_i=x,C_i=c_0}$ and $\E{D_i\mid X_i=x,C_i=c_0}$ are continuous at $x=\bar x$;
    \item[(iii)] $D_i(x,c)=0$ for $x<c$;
    \item[(iv)] $\E{D_i\mid X_i=\bar x,C_i=c_0} >0$.
\end{description}
\end{assumption}

Our target parameter is the following LATE-type parameter:
\begin{align*}
    \tau_{c_0}^{\mathtt{fzy}}(\bar x)
    :=
    \E{Y_i(1)-Y_i(0) \mid X_i=\bar x, C_i=c_0, D_i(\bar x,c_0)=1},
\end{align*}
where $\bar x\in(c_{k^*},c_{k^*+1}]$.
Since $D_i(x,c)\in\{0,1\}$, for $d\in\{0,1\}$ we have
\begin{align*}
    &\E{Y_i(d)D_i(\bar x,c_0)\mid X_i=\bar x,C_i=c_0}\\
    &\quad=
    \E{Y_i(d)\mid X_i=\bar x,C_i=c_0,D_i(\bar x,c_0)=1}
    \E{D_i(\bar x,c_0)\mid X_i=\bar x,C_i=c_0}\\
    &\quad=
    \E{Y_i(d)\mid X_i=\bar x,C_i=c_0,D_i(\bar x,c_0)=1}
    \E{D_i\mid X_i=\bar x,C_i=c_0}.
\end{align*}
It then follows that
\begin{align*}
    \tau_{c_0}^{\mathtt{fzy}}(\bar x)
    =
    \frac{\E{\{Y_i(1)-Y_i(0)\}D_i(\bar x,c_0)\mid X_i=\bar x,C_i=c_0}}{\E{D_i\mid X_i=\bar x,C_i=c_0}}.
\end{align*}
Moreover, the observed conditional mean for group $c_0$ satisfies
\begin{align*}
    \mu_{c_0}(\bar x)
    &=
    \E{Y_i(1)D_i(\bar x,c_0)+Y_i(0)\{1-D_i(\bar x,c_0)\} \mid X_i=\bar x,C_i=c_0}\\
    &=
    \E{\{Y_i(1)-Y_i(0)\}D_i(\bar x,c_0) \mid X_i=\bar x,C_i=c_0}
    + \mu_{0,c_0}(\bar x).
\end{align*}
Therefore,
\begin{align*}
    \E{ \{Y_i(1)-Y_i(0)\}D_i(\bar x,c_0) \mid X_i=\bar x,C_i=c_0}
    =
    \mu_{c_0}(\bar x)-\mu_{0,c_0}(\bar x).
\end{align*}
We thus obtain
\begin{align*}
    \tau_{c_0}^{\mathtt{fzy}}(\bar x)
    =
    \frac{\mu_{c_0}(\bar x)-\mu_{0,c_0}(\bar x) }{\E{D_i\mid X_i=\bar x,C_i=c_0} }.
\end{align*}
The numerator has exactly the same form as the extrapolated treatment-effect parameter in the sharp RD design. The denominator, $p_{c_0}(\bar x):=\E{D_i\mid X_i=\bar x,C_i=c_0}$, is point identified. 
Let $\mathcal B_{\mathtt{num}}(\bar x)$ denote an identified set for $\mu_{c_0}(\bar x)-\mu_{0,c_0}(\bar x)$ obtained by applying the corresponding identification result for the sharp RD design.
Then
\begin{align*}
    \left\{
    \frac{b}{p_{c_0}(\bar x)}
    :
    b\in\mathcal B_{\mathtt{num}}(\bar x)
    \right\}
\end{align*}
gives a valid identified set for $\tau_{c_0}^{\mathtt{fzy}}(\bar x)$.
For example, under Assumptions \ref{assumption:x-monotonicity-OA}(i) and \ref{assumption:c-monotonicity-OA}(i),
\begin{align*}
    \tau_{c_0}^{\mathtt{fzy}}(\bar x) \in \left[
    \dfrac{\mu_{c_0}(\bar x) - \min_{j\in\{k^*+1,\ldots,U\}} \inf_{x \in[\bar x, c_j]} \mu_{c_j}^{-}(x)}{p_{c_0}(\bar x)},\, 
    \dfrac{\mu_{c_0}(\bar x) - \lim_{x\uparrow c_0}\mu_{c_0}(x)}{p_{c_0}(\bar x)}
    \right].
\end{align*}
One can show that the identified set in this example is indeed sharp. 

More generally, however, simply scaling the sharp identified set for the corresponding sharp-RD numerator by $p_{c_0}(\bar x)^{-1}$ need not yield the sharp identified set in the fuzzy-RD design, particularly when restrictions on the support of $Y_i(0)$ are imposed. 
Under one-sided noncompliance, outcomes for untreated individuals above the cutoff directly reveal their untreated potential outcomes and, together with the treatment propensity and support restrictions, impose additional restrictions on the latent untreated potential outcomes of treated individuals. 
Exploiting these restrictions may therefore further tighten the identified set. 

When such additional restrictions arise from support restrictions on the untreated potential outcome, their interaction with smoothness restrictions is also nontrivial, since the fuzzy-RD design can impose additional pointwise restrictions on the counterfactual mean function across values of the running variable.
A detailed analysis of sharp identification in fuzzy RD designs is left for future research.

\subsection{Outline of the Inference Procedure}

The estimation and inference procedures for sharp RD designs extend naturally to the fuzzy RD setting.
The identified set is obtained by scaling the sharp-RD identified set $\mathcal B_{\mathtt{num}}(\bar x)$ by the inverse treatment probability $p_{c_0}(\bar x):=\E{D_i\mid X_i=\bar x,C_i=c_0}$, provided that $p_{c_0}(\bar x)>0$. Since $p_{c_0}(\bar x)$ is consistently estimable, so is the identified set.

To see how inference can be conducted, suppose that the corresponding sharp-RD identified set admits the representation $\mathcal B_{\mathtt{num}}(\bar x)=\{q\in\mathbb R:a q-b\leq G\beta\}$.
Writing $\theta:=\tau_{c_0}^{\mathtt{fzy}}(\bar x)$ and $p:=p_{c_0}(\bar x)$, we have $q=p\theta$, so that the fuzzy-RD identified set can equivalently be written as $\{\theta\in\mathbb R:a p\theta-b\leq G\beta\}$.
Defining the augmented reduced-form parameter $\beta^{\mathtt{fzy}}:=(\beta^\top,p)^\top$, this can be expressed as $\{\theta\in\mathbb R:-b\leq G_\theta\beta^{\mathtt{fzy}}\}$, where $G_\theta:=(G,-a\theta)$.

Thus, for each fixed candidate value $\theta$, the identifying restrictions remain affine in a finite-dimensional vector of point-identified reduced-form parameters, although the loading matrix $G_\theta$ now depends on $\theta$.
The general inference framework of \citet{Andrews_etal:2023} therefore applies after augmenting the reduced-form parameter vector to include the treatment probability and accounting for its sampling variation and covariance with the other estimated reduced-form quantities.
In particular, the covariance matrix of the inequality vector must be recomputed for each candidate value of $\theta$ because $G_\theta$ varies with $\theta$.
As in the sharp RD case, asymptotic validity follows from an appropriate joint Gaussian approximation and consistent covariance estimation for the underlying local-polynomial estimators.

\section{Decision Models for Constant-Bias Assumption}\label{sec:model}

This appendix develops a microeconomic decision model to illustrate when the constant-bias assumption is reasonable and when its validity becomes unclear.
To aid intuition, we present the model in an educational context.

\subsection{Economic Setup}
We begin by describing a model environment tailored to a two-cutoff RD setting.
Building on \citet{Fudenberg_Levine:2022AEJmicro}, we assume that agents decide how much costly effort $e_i$ to invest, influencing their short-term outcome $S_i$ and future outcome $Y_i(0)$ under the control state. 
For example, student $i$ inputs their study effort $e_i$ to achieve a higher test score $S_i$ and higher future earnings $Y_i(0)$. 
In reality, the realized outcomes of $S_i$ and $Y_i(0)$ are not solely determined by the effort and are also affected by stochastic shocks. We model this as follows: $Y_i(0) = y(e_i) + \gamma\mathbf{1}\left\{C_i=h\right\} + \eta_i^y$, and $S_i = s(e_i) + \eta_i^s$, where $y(\cdot)$ and $s(\cdot)$ are structural functions, $\eta_i^y$ and $\eta_i^s$ are zero-mean stochastic errors that are independent of other factors, and $\gamma$ represents a group-level difference in $Y_i(0)$.

Effort incurs a cost $K(e_i, \epsilon_i)$, which also depends on the innate ability $\epsilon_i$, known to the agent $i$. Thus, the cost of a given effort level varies across individuals.

Agents anticipate that they can receive an additional benefit $\tilde{\tau}_i$ in the future if their running variable $X_i$---which may or may not coincide with $S_i$---exceeds a predetermined threshold $C_i$, known to the individual.
This $\tilde{\tau}_i$ represents the agent's subjective belief about the benefit from the treatment, which may differ from the actual one.

In this setup, we assume an agent decides the amount of effort to maximize expected utility:
\begin{align*}
    \max_{e_i} \E{
    v\left(s(e_i) + \eta_i^s\right) - K(e_i, \epsilon_i) +
    \beta\left\{ y(e_i) + \gamma\mathbf{1}\left\{C_i=h\right\} + \eta_i^y + \tilde{\tau}_i \mathbf{1}\left\{i\text{ is treated}\right\} \right\}
    },
\end{align*}
where $v(\cdot)$ represents the utility derived from the short-run outcome $S_i = s(e_i) + \eta_i^s$, $\beta(>0)$ denotes the discount factor, and the expectation is taken with respect to $\eta_i^s$ and $\eta_i^y$. This is simplified as
\begin{align}
    \max_{e_i}\bigg[
    u\left(s(e_i)\right) - K(e_i, \epsilon_i) + \beta\big\{ y(e_i) +\tilde{\tau}_i \P{X_i \geq C_i}
    \big\}
    \bigg],\label{eq: decision model base}
\end{align}
where $u\left(s(e_i)\right)=\E{v\left(s(e_i) + \eta_i^s\right)}$.

\subsection{Decision Problem and the Running Variable Manipulability}
The optimization problem in Equation~\eqref{eq: decision model base} reveals that the agent's decision depends on $\P{X_i \geq C_i}$, the probability of crossing the threshold.
The agent interprets this probability in a different way depending on whether the running variable $X_i$ is influenced by their effort.
Consider two examples:
\begin{itemize}
    \item If financial aid eligibility is based on family wealth level, which is fixed and unaffected by student effort. In this case, $\P{X_i \geq C_i}$ is either 0 or 1, i.e., fully deterministic.

    \item If aid eligibility is determined by a test score, $X_i = S_i$, then the probability of crossing the threshold depends on effort via $s(e_i)$.
\end{itemize}
Following \cite{McCrary:2008}, we say that the running variable is \textit{partially manipulable} when it is under the agent's control but also affected by an idiosyncratic element, i.e., $X_i=S_i=s(e_i)+\eta^s_i$ in our formulation.
In contrast, we say that the running variable is \textit{non-manipulable} when it is not a function of the agent's effort.

Agents face one of two decision problems depending on whether the running variable is manipulable:
\begin{subequations}\label{eq: decision models}
    \begin{align}
        &\max_{e_i}\bigg[
        u\left(s(e_i)\right) - K(e_i, \epsilon_i) + \beta y(e_i) 
        \bigg]\text{ if }X_i\text{ is non-manipulable},\label{eq: decision models W}\\
        &\max_{e_i}\bigg[
        u\left(s(e_i)\right) - K(e_i, \epsilon_i) + \beta\big\{ y(e_i) +\tilde{\tau}_i \P{s(e_i) + \eta_i^s \geq C_i}
        \big\}
        \bigg]\text{ if }X_i=S_i.\label{eq: decision models S}
    \end{align}
\end{subequations}
We will write the optimal level of effort by $e_i^*$.

\subsection{Non-Manipulable Running Variable Case}
We are now in a position to analyze the constant-bias assumption within our model environment. We begin with the case where the running variable is non-manipulable, corresponding to equation \eqref{eq: decision models W}.
An immediate implication from \eqref{eq: decision models W} is that the optimal effort level $e_i^*$ does not depend on the cutoff $C_i$. 
This means that differences across groups can only arise from differences in the distribution of ability $\epsilon_i$ (through the cost function) and from the group-level shift $\gamma$ in outcomes.
This leads to the following result:
\begin{prop}\label{prop: characterization of constant-bias, similar group case}
    Suppose that the running variable is non-manipulable.
    If $\epsilon_i \indep C_i \mid X_i$, then the constant-bias assumption holds.
\end{prop}
\begin{remark}
    This is not the if-and-only-if result. Instead, it provides an interpretable sufficient condition for the constant-bias assumption.
\end{remark}

Proposition \ref{prop: characterization of constant-bias, similar group case} says that the constant-bias assumption holds if cutoff assignment is as good as random conditional on the running variable.
This interpretation is appealing for empirical researchers, as it links the validity of the constant-bias assumption to a familiar random assignment-like condition often invoked in the causal inference literature.

In this view, the proposition offers a useful reference point for assessing whether the constant-bias assumption is reasonable in practical applications.
If an economist believes that the cutoff positions are set ``exogenously" and that groups are comparable in terms of unobserved characteristics, then the constant-bias assumption is likely to hold in settings with a non-manipulable running variable.
In contrast, the assumption can fail when the cutoff is correlated with unobserved group characteristics.

\subsection{Partially Manipulable Running Variable Case}\label{subsubsec: manipulable}
We now turn to the case where the running variable is partially manipulable, as characterized by equation \eqref{eq: decision models S}.
To proceed, we impose some conditions. 
\begin{assumption}\label{assumption: regularity conditions} 
\begin{itemize}
    \item[(i)] $u\left(s(e_i)\right) + \beta\left\{y(e_i) + \tilde{\tau}_i\P{s(e_i) + \eta_i^s \geq C_i}\right\}$ is strictly concave in $e_i$, 
    \item[(ii)] $u\left(s(e_i)\right)$, $K(e_i, \epsilon_i)$, $y(e_i)$, $s(e_i)$, and $\P{s(e_i) + \eta_i^s \geq C_i}$ are continuously differentiable in $e_i$. Furthermore, $s^\prime(e_i)>0$ and $K(e_i, \epsilon_i)$ is continuous in $(e_i, \epsilon_i)$.
    \item[(iii)] The distribution of the shock $\eta_i^s$ has a density function $f_{\eta^s}$. 
    \item[(iv)] $K(e_i, \epsilon_i)$ is convex in $e_i$,
    \item[(v)] The support of $\epsilon_i$ is an interval and identical across both groups. Furthermore, $e_i\in[0, \overline{e}]$.
\end{itemize}
\end{assumption}
\begin{assumption}\label{assumption: homo tau}
    $\tilde{\tau}_i = \tilde{\tau}>0$ for every $i$.
\end{assumption}
Assumption~\ref{assumption: regularity conditions} comprises regularity conditions. 
Assumptions \ref{assumption: regularity conditions}(i) and (ii) are smoothness and (high-level) concavity assumptions that ensure analytical traceability. 
Assumption \ref{assumption: regularity conditions}(iii) is also a weak smoothness assumption.
Assumption \ref{assumption: regularity conditions}(iv) requires convexity of the cost function, covering commonly used specifications such as the linear and quadratic cost functions. Assumption \ref{assumption: regularity conditions}(v) imposes mild support conditions.

Assumption~\ref{assumption: homo tau} rules out heterogeneity in agents’ beliefs about treatment benefits. While strong, the next result suggests that \textit{even under this homogeneity assumption}, the constant-bias assumption may not hold.

\begin{prop}\label{prop: dependence of optimal effort}
    Suppose that the running variable is partially manipulable and that Assumptions \ref{assumption: regularity conditions}-\ref{assumption: homo tau} are satisfied. 
    Suppose also that the solution $e_i^*$ is an interior point of $[0, \overline{e}]$ for any $\epsilon_i$.
    Then the optimal effort $e_i^*$ does not depend on $C_i$ for any $\epsilon_i$ if and only if the density function $f_{\eta^s}$ is periodic with period $h-\ell$, i.e., $f_{\eta^s}(z)=f_{\eta^s}(z+h-\ell)$, on the interval for every $z$ such that $z, z+(h-\ell) \in [\ell-\sup_{\epsilon}s(e^*(\epsilon)), h-\inf_{\epsilon}s(e^*(\epsilon))]$.
\end{prop}

In most empirical settings, the condition made on the density function is questionable.
We typically have that $l < \sup_{\epsilon}s(e^*(\epsilon))$ and $\inf_{\epsilon}s(e^*(\epsilon)) < h$, so the interval $[l-\sup_{\epsilon}s(e^*(\epsilon)), h-\inf_{\epsilon}s(e^*(\epsilon))]$ includes zero.
Now, recalling that $f_{\eta^s}$ represents the distribution of idiosyncratic noise in the score, it is far more natural to assume that this density is unimodal and centered at zero, such as the Gaussian errors, rather than being periodic. 
Therefore, the proposition states that the optimal effort generally \textit{does} depend on the cutoff position $C_i$.

The dependence of the effort input $e^*_i$ on $C_i$ implies that the distribution of $S_i$ and $Y_i(0)$, which are functions of $e^*_i$, can differ between the two groups.
Hence, in contrast to the case with a non-manipulable running variable, the validity of the constant-bias assumption is not guaranteed, even under random assignment, identical distributions of $\epsilon_i$, and the strong homogeneity assumption $\tilde{\tau}_i = \tilde{\tau}$.
Of course, the dependence of $e_i^*$ on $C_i$ does not rule out the possibility that the constant-bias assumption holds; however, motivating this assumption in practice may be challenging, since its validity depends on the functional forms of the structural functions, which are not observable by economists.

\subsection{Proofs for Appendix \ref{sec:model}}

\begin{proof}[Proof of Proposition \ref{prop: characterization of constant-bias, similar group case}]
    Under the non-manipulable running-variable case, the objective function in \eqref{eq: decision models W} does not depend on $C_i$. Hence, the optimal effort can be written as $e_i^*=e^*(\epsilon_i)$, independently of $C_i$. It follows that, for $c\in\{l,h\}$,
    \begin{align*}
        \mu_{0,c}(x)
        =
        \E{Y_i(0)\mid X_i=x,C_i=c}
        =
        \E{y(e^*(\epsilon_i))\mid X_i=x,C_i=c}
        +\gamma\mathbf{1}\{c=h\},
    \end{align*}
    where we use the independence and zero-mean assumptions on $\eta_i^y$.
    Since $\epsilon_i\indep C_i\mid X_i$, the first term is identical across the two cutoff groups conditional on $X_i=x$. Therefore,
    \begin{align*}
        \mu_{0,h}(x)-\mu_{0,l}(x)=\gamma,
    \end{align*}
    which is constant in $x$. Hence, the constant-bias assumption holds.
\end{proof}

\begin{proof}[Proof of Proposition \ref{prop: dependence of optimal effort}]
    Let $e_c^*(\epsilon_i)$ denote the optimal effort of those with $\epsilon_i$ in the group $c\in\{l,h\}$. Take an arbitrary $\epsilon_i$ and suppose $e_l^*(\epsilon_i)=e_h^*(\epsilon_i)=: e_i^*$. Then, by the first-order condition, we have
    \begin{align*}
        &u^\prime\left(s(e_i^*)\right)s^\prime(e_i^*) - \frac{\partial K}{\partial e}(e_i^*, \epsilon_i) + \beta(y^\prime(e_i^*) + \tilde{\tau}s^\prime(e_i^*)f_{\eta^s}(l-s(e_i^*)))\\
        &\quad= u^\prime(s(e_i^*)s^\prime(e_i^*) - \frac{\partial K}{\partial e}(e_i^*, \epsilon_i) + \beta(y^\prime(e_i^*) + \tilde{\tau}s^\prime(e_i^*)f_{\eta^s}(h-s(e_i^*))) = 0.
    \end{align*}
    This implies $f_{\eta^s}(l-s(e_i^*))=f_{\eta^s}(h-s(e_i^*))$.
    The objective function in \eqref{eq: decision models S} is continuous in $(e,\epsilon)$ and the choice set $[0,\overline e]$ is compact. Hence, by Berge's maximum theorem, the argmax correspondence is upper hemicontinuous and nonempty. Moreover, by strict concavity in $e$, the maximizer is unique, so the optimal effort $e_c^*(\epsilon_i)$ is continuous in $\epsilon_i$. Since $e_l^*(\epsilon_i)=e_h^*(\epsilon_i)=: e_i^*$ by assumption, the map $\epsilon_i\mapsto s(e^*(\epsilon_i))$ is continuous.
    Then, recalling that $f_{\eta^s}(l-s(e_i^*))=f_{\eta^s}(h-s(e_i^*))$, $f_{\eta^s}$ must be periodic with period $h-l$ on the interval $[l-b, h-a]$, where $a=\inf_{\epsilon}s(e^*(\epsilon))$ and $b=\sup_{\epsilon}s(e^*(\epsilon))$.

    Conversely, let $a_\ell := \inf_{\epsilon}s(e_\ell^*(\epsilon))$ and $b_\ell := \sup_{\epsilon}s(e_\ell^*(\epsilon))$, and suppose that
    \begin{align*}
        f_{\eta^s}(z) = f_{\eta^s}(z+h-\ell)
        \qquad \text{for every } z\in [\ell-b_\ell,h-a_\ell].
    \end{align*}
    Take an arbitrary $\epsilon_i$. 
    By definition, $\ell-s(e_\ell^*(\epsilon)) \in [\ell-b_\ell,\ell-a_\ell]$ and $h-s(e_\ell^*(\epsilon)) \in [h-b_\ell,h-a_\ell]$.
    Therefore, both $\ell-s(e_\ell^*(\epsilon))$ and $h-s(e_\ell^*(\epsilon))$ belong to $[\ell-b_\ell,h-a_\ell]$, and the periodicity condition gives $f_{\eta^s} ( \ell-s(e_\ell^*(\epsilon)) ) = f_{\eta^s} ( h-s(e_\ell^*(\epsilon)) )$.
    
    Since $e_l^*(\epsilon_i)$ is an interior optimum under $C_i=l$, it satisfies
    \begin{align*}
        &u^\prime\left(s(e_l^*(\epsilon_i))\right)s^\prime(e_l^*(\epsilon_i))
        -
        \frac{\partial K}{\partial e}\left(e_l^*(\epsilon_i),\epsilon_i\right)\\
        &\qquad
        +
        \beta \left[y^\prime(e_l^*(\epsilon_i)) + \tilde{\tau} s^\prime(e_l^*(\epsilon_i)) f_{\eta^s} \left(l-s(e_l^*(\epsilon_i))\right)\right]
        =0.
    \end{align*}
    Substituting $f_{\eta^s} ( \ell-s(e_\ell^*(\epsilon)) ) = f_{\eta^s} ( h-s(e_\ell^*(\epsilon)) )$ into the preceding first-order condition yields
    \begin{align*}
        &u^\prime\left(s(e_l^*(\epsilon_i))\right)s^\prime(e_l^*(\epsilon_i))
        -
        \frac{\partial K}{\partial e}\left(e_l^*(\epsilon_i),\epsilon_i\right)\\
        &\qquad
        +
        \beta \left[y^\prime(e_l^*(\epsilon_i)) + \tilde{\tau} s^\prime(e_l^*(\epsilon_i)) f_{\eta^s} \left(h-s(e_l^*(\epsilon_i))\right)\right]
        =0.
    \end{align*}
    Thus, $e_l^*(\epsilon_i)$ also satisfies the first-order condition for the optimization problem under $C_i=h$. By strict concavity of the objective function, the optimizer is unique, and therefore $e_h^*(\epsilon_i)=e_l^*(\epsilon_i)$.
    Since $\epsilon_i$ was arbitrary, the equality holds for all $\epsilon_i \in \mathrm{supp}(\epsilon_i)$. This proves the converse.
\end{proof}

\end{document}